\documentclass[twocolumn]{emulateapj}
\slugcomment{{\sc Accepted to ApJ:} June 7, 2013} 
\usepackage{amsmath}
\usepackage{amssymb}
\usepackage{graphicx}
\usepackage{subfigure}

\newcommand{\ergcms}{erg$\,$cm$^{-2}$$\,$s$^{-1}$}
\newcommand{\ergs}{erg$\,$s$^{-1}$}

\begin{document}
   \title{Transit observations of the Hot Jupiter HD~189733\lowercase{b} at X-ray wavelengths}

\author{K.\ Poppenhaeger\altaffilmark{1,2}, J.\ H.\ M.\ M.\ Schmitt\altaffilmark{2}, S.\ J.\ Wolk\altaffilmark{1}}
\affil{\altaffilmark{1}Harvard-Smithsonian Center for Astrophysics, 60 Garden Street, Cambridge, MA 02138, USA\\
\altaffilmark{2}Hamburger Sternwarte, Gojenbergsweg 112, 21029 Hamburg, Germany}
\email{kpoppenhaeger@cfa.harvard.edu}

% \abstract{}{}{}{}{} 
% 5 {} token are mandatory
 
\begin{abstract}
We present new X-ray observations obtained with {\it Chandra} ACIS-S of the HD~189733 system, consisting of a K-type star orbited by a transiting Hot Jupiter and an M-type stellar companion. We report a detection of the planetary transit in soft X-rays with a significantly larger transit depth than observed in the optical. The X-ray data favor a transit depth of 6-8\%, versus a broadband optical transit depth of 2.41\%. While we are able to exclude several possible stellar origins for this deep transit, additional observations will be necessary to fully exclude the possibility that coronal inhomogeneities influence the result. From the available data, we interpret the deep X-ray transit to be caused by a thin outer planetary atmosphere which is transparent at optical wavelengths, but dense enough to be opaque to X-rays. The X-ray radius appears to be larger than the radius observed at far-UV wavelengths, most likely due to high temperatures in the outer atmosphere at which hydrogen is mostly ionized. We furthermore detect the stellar companion HD~189733B in X-rays for the first time with an X-ray luminosity of $\log L_X = 26.67$\,\ergs. We show that the magnetic activity level of the companion is at odds with the activity level observed for the planet-hosting primary. The discrepancy may be caused by tidal interaction between the Hot Jupiter and its host star.
\end{abstract}


%________________________________________________________________

\section{Introduction}

Since the first exoplanet detections almost two decades ago, the research focus of the field has widened from merely finding exoplanets to characterizing and understanding their physical properties. Many exoplanets are found in very close orbits around their host stars -- unlike planets in the solar system -- and are therefore subject to strong stellar irradiation. Theoretical models predict that the incident stellar flux can deposit enough energy in the planetary atmosphere to lift parts of it out of the planet's gravitational well, so that the planetary atmosphere evaporates over time. This evaporation process is thought to be driven by X-ray and extreme-UV irradiation, and different theoretical models emphasize aspects such as Roche lobe effects or hydrodynamic blow-off conditions \citep{Lecavelier2004, Lammer2003, Erkaev2007}. First observational evidence for extended planetary atmospheres and possibly evaporation was found for the Hot Jupiters HD~209458b and HD~189733b, some observations showing temporal variations in the escape rate \citep{Vidal-Madjar2003, Lecavelier2010, Lecavelier2012}. The lower limit of the mass loss, inferred from absorption in H\,{\sc i} Lyman-$\alpha$ during transit, was determined to be on the order of $\gtrsim 10^{10}$~g\,s$^{-1}$, but escape rates may well be several orders of magnitude larger \citep{Lecavelier2004}. 
  
Here we investigate the HD~189733 system, located at a distance of 19.45\,pc from the Sun. It consists of two stellar components and one known extrasolar planet: The planet-hosting primary HD~189733A is a main-sequence star of spectral type K0 and is orbited by a transiting Hot Jupiter, HD~189733b, with $M_p = 1.138 M_{Jup}$ in a 2.22\,d orbit (see Table~\ref{basic_properties} for the properties of the star-planet system). There is a nearby M4 dwarf located at $11.4^{\prime\prime}$ angular distance, which was shown to be a physical companion to HD~189733A in a 3200\,yr orbit based on astrometry, radial velocity and common proper motion \citep{Bakos2006}.

HD~189733b is one of the prime targets for planetary atmosphere studies, as it is the closest known transiting Hot Jupiter. Different molecules and atomic species have been detected in its atmosphere with transmission spectroscopy, see for example \citep{Tinetti2007, Redfield2008}. Recent transit observations at UV wavelengths indicate that the transit depth in the H\,{\sc i} Lyman-$\alpha$ line might be larger than in the optical \citep{Lecavelier2010, Lecavelier2012}, possibly indicating the presence of extended planetary atmosphere layers beyond the optical radius. To test this hypothesis, we repeatedly observed the HD~189733 system in X-rays during planetary transits.

\begin{table}[t!]
\begin{center}
\begin{tabular}{l l}
\hline \hline
stellar parameters:		& 	\\ \hline
spectral type			& K0V	\\
mass ($M_\odot$)		& 0.8	\\
radius ($R_\odot$)		& 0.788 \\ 
distance to system (pc)		& 19.3 \\ \hline
planetary parameters:		& 	\\ \hline
mass $\times \sin i$ ($M_{Jup}$)& 1.138	\\
optical radius ($R_{Jup}$)	& 1.138	\\
semimajor axis (AU)		& 0.03142		\\
orbital period (d)		& 2.21857312		\\ 
mid-transit time (BJD)		& 2453988.80261 	\\ 
orbital inclination		& 85.51$^\circ$\\ \hline
\end{tabular}
\caption{Properties of the HD~189733~Ab system, as given in the Extrasolar Planet Encyclopedia (www.exoplanet.eu).}
\label{basic_properties}
\end{center}
\end{table}

%__________________________________________________________________

\section{Observations and data analysis}

We observed the HD~189733 system in X-rays during six planetary transits with {\it Chandra} ACIS-S. One additional X-ray transit observation is available from {\it XMM-Newton} EPIC. Another observation was conducted with {\it Swift} in 2011; however, this observation has much lower signal to noise and does not cover the whole transit. 

The {\it Chandra} observations all lasted $\approx 20$\,ks, the {\it XMM-Newton} pointing has a duration of $\approx 50$\,ks, and all data sets are approximately centered on the transit (see Table~\ref{obslog} for  details of the observations). The orbital phase coverage is therefore highest for phases near the transit, with the transit lasting from $0.984-1.016$ (first to fourth contact); the phase coverage quickly decreases for smaller or larger phases until coverage is only provided by the single {\it XMM-Newton} observation (see Fig.~\ref{phase}). {\it Chandra} uses a dithering mode; this does not produce signal changes in the observations 1, 2, 3, 4, and 6. In observation 5, the source was dithered near a bad pixel column, resulting in the loss of a few photons every 1000\,s. The transit lasts ca.\ 6500\,s and the fraction of missing photons is slightly larger out of transit than during transit, so that no spurious transit signal or limb effects can be produced by this.

\begin{table}[t!]
\begin{footnotesize}
\begin{tabular}{l l l l}
\hline \hline
Instrument		& ObsID		& start time		& duration 	\\ 
			&		&			& (ks)	\\
\hline
XMM-Newton EPIC		& 0506070201	& 2007-04-17 14:06:57	& 54.5	\\
Chandra ACIS-S		& 12340		& 2011-07-05 11:56:51	& 19.2	\\
Chandra ACIS-S		& 12341		& 2011-07-21 00:23:17	& 19.8	\\
Chandra ACIS-S		& 12342		& 2011-07-23 06:18:51	& 19.8	\\
Chandra ACIS-S		& 12343		& 2011-07-12 03:26:19	& 19.8	\\
Chandra ACIS-S		& 12344		& 2011-07-16 13:45:59	& 18.0	\\
Chandra ACIS-S		& 12345		& 2011-07-18 19:08:58	& 19.8	\\
\hline
\end{tabular}
\caption{X-ray observation details for the HD~189733 system.}
\label{obslog}
\end{footnotesize}
\end{table}

\subsection{Spatially resolved X-ray images of the HD~189733 system}

   \begin{figure}[t!]
   \centering
   \includegraphics[width=0.45\textwidth]{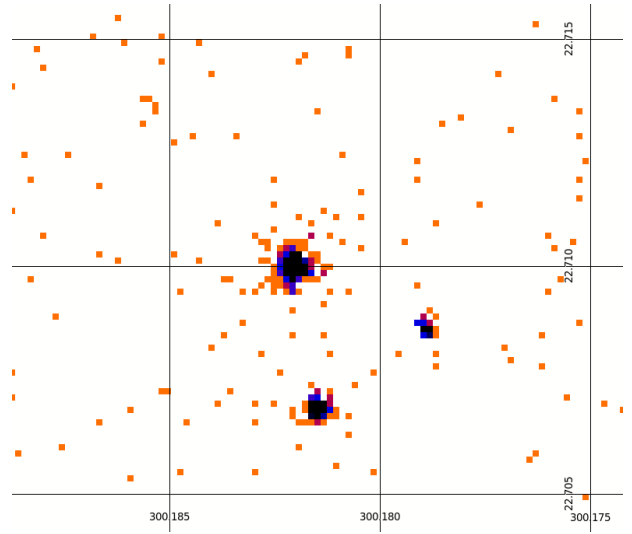}
   \caption{X-ray image of the planet-hosting star HD~189733A (upper left object), its companion star HD~189733B (right object) and a background object (lower object). The image is extracted from a single {\it Chandra} exposure with an integration time of 20\,ks in the 0.1-5\,keV energy band. The companion HD~189733B is detected for the first time in X-rays with these observations.}
   \label{Skypic}%
   \end{figure}

All previous obervations of the HD~189733 system had been performed with {\it XMM-Newton}; due to the rather broad PSF of the prime instrument PN (ca.\ $12.5^{\prime\prime}$ FWHM), the secondary HD~189733B was not properly resolved in those observations. Our {\it Chandra} observations with a PSF of ca.\ $0.5^{\prime\prime}$ (FWHM) fully resolve the stellar components for the first time; a soft X-ray image extracted from one single {\it Chandra} exposure with 20\,ks duration is shown in Fig.~\ref{Skypic}. The primary HD~189733A is shown in the center, the secondary HD~189733B to the lower right at an angular distance of $11.4^{\prime\prime}$ to the primary, and a background source at the bottom of the image at an angular distance of $11.7^{\prime\prime}$. The planetary orbit of HD~189733b is not resolved in these observations: if observed at quadrature, its angular distance to the host star would be ca.\ $1.6$\,mas.

\subsection{Individual light curves}

To inspect the individual X-ray light curves of HD~189733A, HD~189733B, and the background source, we extracted X-ray photons from an extraction region centered on each source with $2.5^{\prime\prime}$ radius for the {\it Chandra} observations. We restricted the energy range to 0.1-2.0\,keV for HD~189733A and HD~189733B, because practically no stellar flux at higher energies was present; we used an energy band of 0.1-10.0\,keV for the background source. For the archival {\it XMM-Newton} observation, we chose an extraction radius $10^{\prime\prime}$ to collect most of the source photons from HD~189733A while avoiding strong contamination from the other two sources. {\it XMM-Newton} EPIC/PN provides sensitivity down to 0.2\,keV, so we restricted the energy range to 0.2-2.0\,keV. In all {\it Chandra} observations the background signal was neglibigly low and constant (0.05\% of the source count rate). For {\it XMM-Newton}, the background signal is weak and only slightly variable during times that are interesting for the transit analysis (ca.\ 2\% of the source count rate); however, in the beginning of the observation the background is stronger, and we therefore followed the standard procedure and extracted photons from a large source-free area, scaled down this background signal to the source extraction area and subtracted the background signal from the {\it XMM-Newton} source light curve.

For the light curves, we chose time bin sizes suited to obtain reasonable error bars; specifically, the chosen bin size is 200\,s for HD~189733A, 1000\,s for HD~189733B, and 500\,s for the background source.

\subsection{Phase-folded light curve}\label{phasedlc}

To test if a transit signal can be detected in the X-ray data, we also constructed a phase-folded, added-up X-ray light curve. We corrected all X-ray photon arrival times with respect to the solar system barycenter, using the ciao4.2 task {\it axbary} for the {\it Chandra} events and the SASv11.0 task {\it barycen} for the {\it XMM-Newton} events. Orbital phases were calculated using the mid-transit time and orbital period given in the Extrasolar Planet Encyclopaedia and are listed in Table~\ref{basic_properties}. We first extracted light curves from the individual observations with a very small time binning (10\,s). We normalized the light curves to an out-of-transit level of unity, chose the desired phase segments (0.005, corresponding to roughly 16 minutes), and added up the individual observations to form the binned phase-folded lightcurve, giving the same weight to each individual light curve. These 0.005 phase bins contain $\approx 500$ X-ray photons each. 
The error bars of the individual light curves are dominated by Poissonian noise, which is close to Gaussian noise given the number of X-ray counts per 0.005 phase bin; these errors were propagated to the added-up light curve.
Because of the intrinsic variability of the stellar corona it is important to average over the individual transit observations. We therefore restrict our analysis to orbital phases at which at least six of the seven observations provide signal. Specifically, this is given at orbital phases between $0.9479$ to $1.0470$.

   \begin{figure}[t!]
   \centering
   \includegraphics[width=0.5\textwidth]{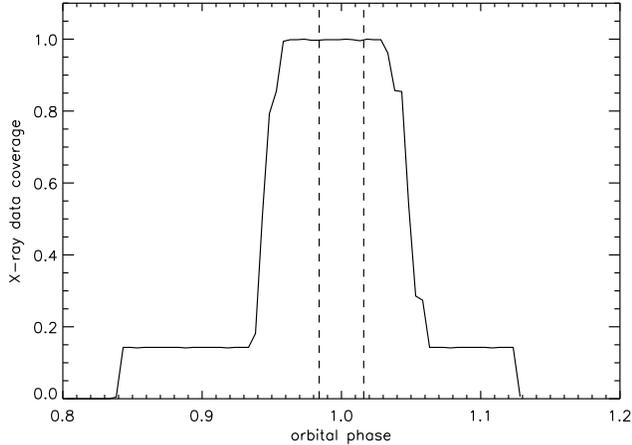}
   \caption{Cumulative phase coverage of the available X-ray data for HD~189733A near the planetary transit. The added exposures are normalized to unity; at this maximum value, a given phase is covered by seven individual X-ray observations. The duration of the optical transit (first to fourth contact) is depicted by dashed vertical lines.}
   \vspace{0.5cm}
   \label{phase}%
   \end{figure}

\subsection{Spectra}

We extracted CCD spectra of all three sources from the {\it Chandra} ACIS-S exposures. We used a minimum binning of 15 counts per energy bin; for the secondary, we used smaller bins with $\geq 5$ counts per bin because of the low count rate. We fitted the resulting six spectra of each source simultaneously in Xpsec~v12.0, therefore ignoring spectral changes inbetween the individual observations and focusing on the overall spectral properties of each object. We fit the spectra with thermal plasma models for HD~189733A and HD~189733B; since we do not know the exact nature of the third source {\it a priori}, we test a thermal plasma model and a power law model for fitting those spectra. We use elemental abundances from \cite{GrevesseSauval1998} for all spectral fits.

A detailed analysis of HD~189773A's spectrum observed with {\it XMM-Newton} (out of transit) was performed before; see \cite{Pillitteri2010}, \cite{Pillitteri2011}.

\section{Results}

\subsection{HD~189733A: Individual light curves and spectra}\label{indivlcs}

   \begin{figure}[t!]
   \centering
   \includegraphics[width=0.5\textwidth]{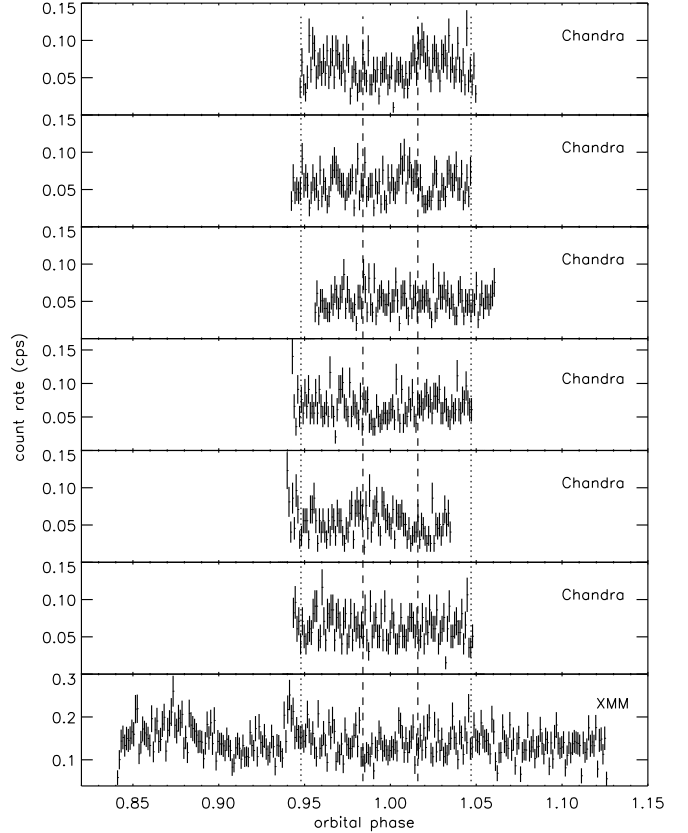}
   \caption{Individual X-ray light curves of HD~189733A which cover the planetary transit, with a time binning of 200\,s and $1\sigma$ Poissonian error bars. All light curves are extracted from an energy range of 0.2-2\,keV.}
   \label{individual_lcs}%
   \end{figure}
   
   \begin{figure}[t!]
   \centering
   \includegraphics[width=0.405\textwidth]{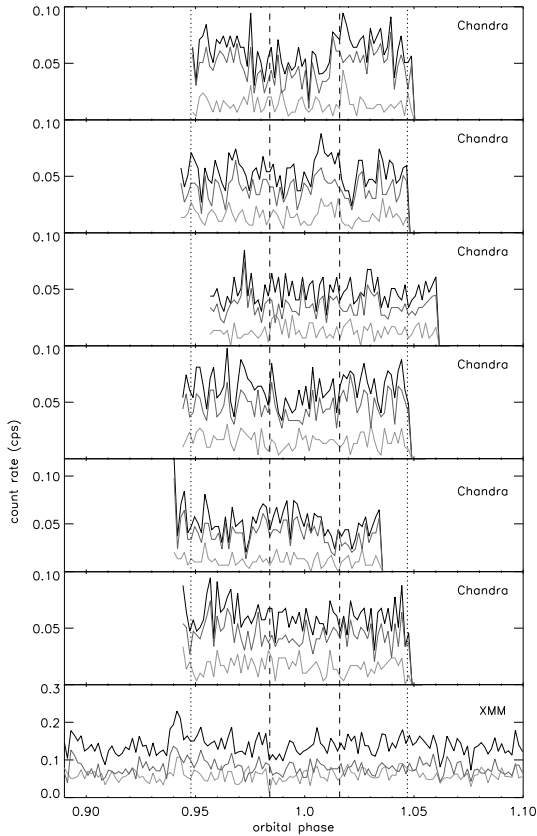} 
   \caption{The individual X-ray light curves of HD~189733A in different energy bands (light grey: 0.2-0.6\,keV, grey: 0.6-2.0\,keV, black: full energy band), with a time binning of 300\,s.}
   \label{XrayLC_hard_soft}%
   \end{figure}

The individual X-ray light curves of HD~189733A are shown in Fig.~\ref{individual_lcs}. The mean count rate is at $\approx 0.07$\,cts\,s$^{-1}$ for {\it Chandra}'s ACIS-S, and at $\approx 0.15$\,cts\,s$^{-1}$ for {\it XMM-Newton}'s EPIC. 

An inspection of these light curves shows their intrinsic variability to be quite small for an active star with maximum count rate variations being less than a factor of two. Such small changes are often found to be the base level of variability in cool stars, and are referred to as ''quasi-quienscence'' \citep{RobradePoppenhaeger2010}. Larger flares can be superimposed on this type of variability, but this is not the case for our observations of HD~189733A. 

However, small flare-shaped variations are present, for example in the second {\it Chandra} light curve at orbital phase 1.01. For the X-ray flares of HD~189733 which have been detected with {\it XMM-Newton} at other orbital phases, \cite{Pillitteri2010, Pillitteri2011} show that changes in the hardness ratio do not always accompany the flares, so that we focus on the light curves as a primary flare indicator for our observations. We show the individual light curves split up into a hard and soft band (0.2-0.6\,keV, 0.6-2\,keV) together with the total band in Fig.~\ref{XrayLC_hard_soft}, using a bin size of 300~s to have reasonable photon statistics for the individual bands. We test for correlated intensity changes in both bands in detail in section \ref{lcselection}, finding that the second {\it Chandra} observation and the {XMM-Newton} observation likely contain small stellar flares.

\begin{table}[t!]
\begin{tabular}{l l l}
\hline \hline
object				& HD 189733A			& HD 189733B	\\ \hline
$T_1$ (keV)			& $0.25\pm0.01$			& $0.32\pm0.02$		\\
$norm_1$			& $(1.81\pm0.1)\times 10^{-4}$	& $(5.1\pm0.5)\times 10^{-6}$	\\
$EM_1$ ($10^{50}$\,cm$^{-3}$)	& $7.7\pm0.4$			& $0.22\pm0.02$	\\
$T_2$ (keV)			& $0.62\pm0.03$			& $1.25\pm0.5$	\\
$norm_2$			& $(6.6\pm0.5)\times 10^{-5}$	& $(1.7\pm0.5)\times 10^{-6}$	\\
$EM_2$ ($10^{50}$\,cm$^{-3}$)	& $2.8\pm0.2$			& $0.07\pm0.02$	\\
O				& $0.31\pm0.02$			& (fixed at solar)		\\
Ne				& $0.25\pm0.13$			& (fixed at solar)		\\ \vspace{0.2cm}
Fe				& $0.64\pm0.05$			& (fixed at solar)		\\ \vspace{0.2cm}
$\chi^2_{red}$ (d.o.f.)		& 1.43 (279)			& 1.12 (64)		\\
$F_X$ (0.25-2 keV)		& $2.63\times 10^{-13}$  	& $1.1\times 10^{-14}$		\\
$\log L_X$ (0.25-2 keV)		& 28.1				& 26.67				\\ \hline
\end{tabular}
\caption{Thermal plasma models for the X-ray emission of the primary and secondary star. Flux is in units of erg\,s$^{-1}$\,cm$^{-2}$, luminosity in units of erg\,s$^{-1}$. The emission measure is derived from Xspec's fitted normalization by multiplying it with $4\pi\,d^2\times10^{14}$, with $d$ being the distance to the two stars.}
\label{spec_AB}
\end{table}

\begin{figure}[t!]
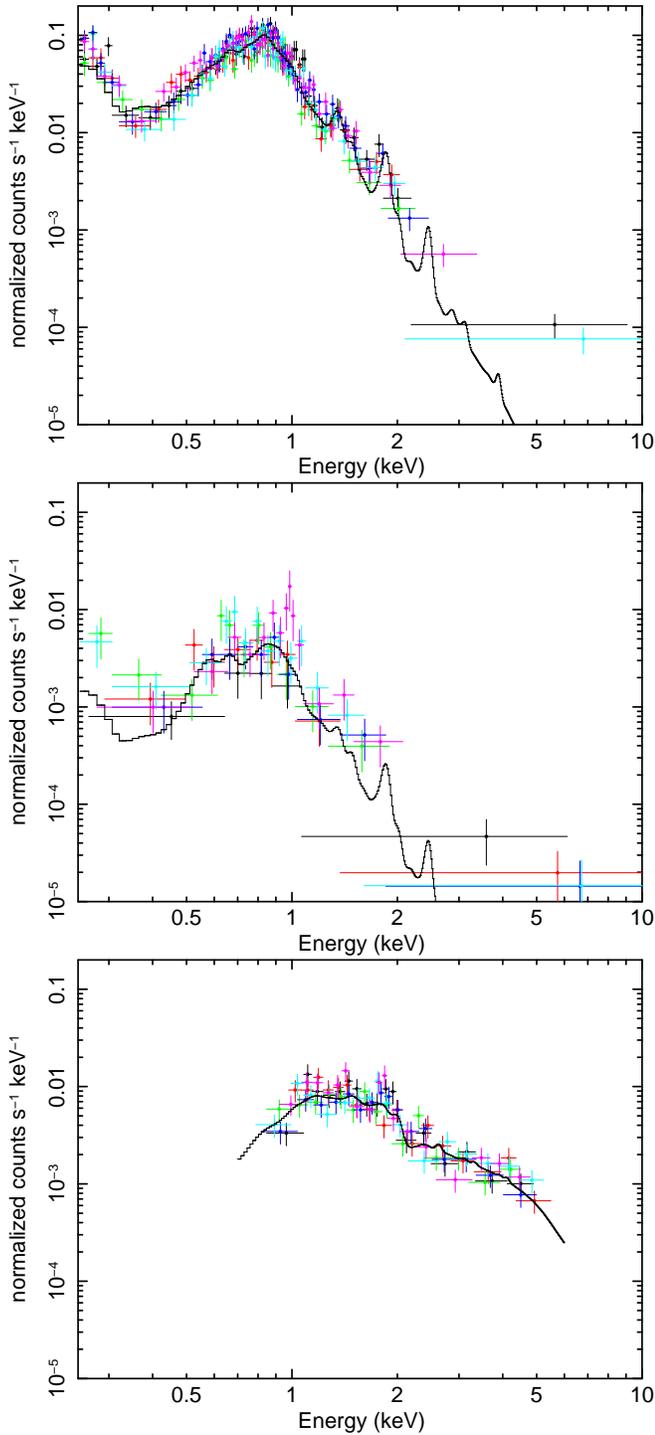

   \centering
   \includegraphics[width=0.35\textwidth,angle=-90]{prim.eps}
   \includegraphics[width=0.35\textwidth,angle=-90]{sec.eps}
   \includegraphics[width=0.35\textwidth,angle=-90]{unk.eps}
   \caption{X-ray spectra of HD~189733A (top panel), its stellar companion HD~189733B (middle panel), and the third nearby X-ray source (lower panel). The black lines represents the respective fitted models listed in Tables~\ref{spec_AB} and \ref{spec_unk}. In the online color version of this figure, different colors indicate the different observations.}
   \label{spectra}%
   \end{figure}

The source spectra of HD~189733A, extracted from the six {\it Chandra} ACIS-S pointings, are shown in Fig.~\ref{spectra}, top panel. The dominant part of the emission is at energies below 2\,keV, and all six observations display very similar energetic distributions and intensities. We fitted the spectra simultaneously with a thermal plasma model; two temperature components and variable abundances for the most prominent elements at soft X-ray energies (oxygen, neon, and iron) are necessary for a decent fit quality. We find a mean coronal temperature of ca. 4.5\,MK, with a dominant temperature component at ca. 3\,MK (see Table~\ref{spec_AB}). Our spectral fit yields an X-ray flux of $2.63\times 10^{-13}$\,\ergcms \ in the 0.25-2\,keV energy band, corresponding to an X-ray luminosity of $L_X = 1.1\times 10^{28}$\,\ergs. We calculate HD~189733A's bolometric luminosity, using bolometric corrections from \cite{Flower1996}, to be $L_{bol} = 0.35\times L_{bol,\,\odot} = 1.3\times 10^{33}$\,\ergs. The ratio of X-ray to bolometric luminosity is therefore $R_X = L_X/L_{bol} = 10^{-5}$, characterizing HD~189733A as a moderately active star; the Sun, for comparison, emits varying X-ray flux fractions of about $10^{-6} - 10^{-7}$ during its activity cycle. 

The coronal abundances show a FIP-bias, meaning that the abundances of elements with a high {\bf f}irst {\bf i}onization {\bf p}otential (oxygen and neon) are lower than the abundances of low-FIP elements (iron); all abundances are given relative to solar photospheric values \citep{GrevesseSauval1998}. This effect is observed in coronae of stars with low to moderate activity, while high-activity stars show an inverse FIP-effect. Recently, a possible correlation of FIP-bias with spectral type has been found for inactive and moderately active stars \citep{WoodLinsky2010}. Our measured abundances for HD~189733A fit into this pattern: The average logarithmic abundance of neon and oxygen relative to iron is $-0.41$ here, and \cite{WoodLinsky2010} predict similar values around $-0.35$ for stars of spectral type K0.

\subsection{HD~189733B}

\begin{figure}[h!]
   \centering
   \includegraphics[width=0.5\textwidth]{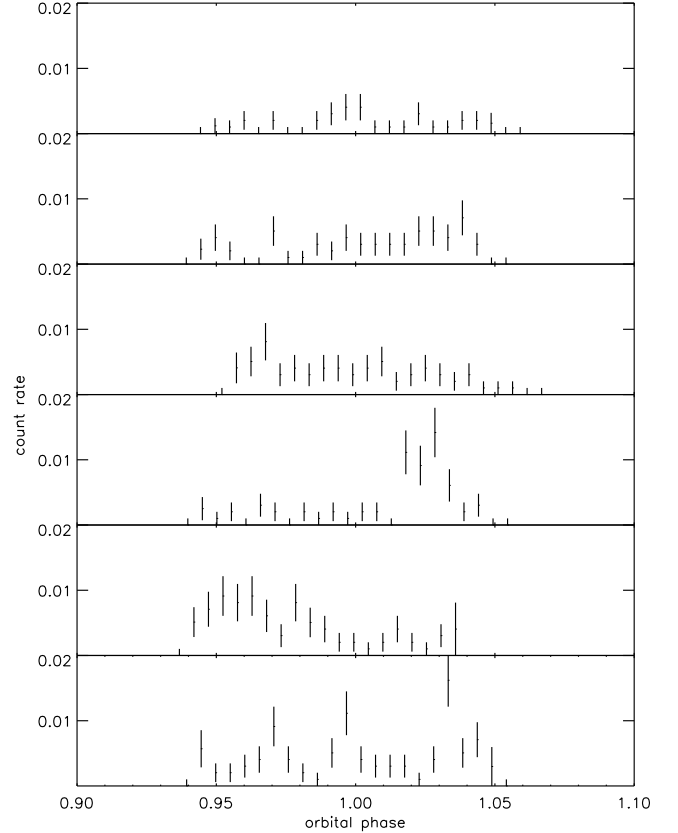}
   \caption{Individual X-ray light curves of HD~189733B, taken with {\it Chandra} ACIS-S (the source is unresolved in {\it XMM-Newton}). All light curves are extracted from an energy range of 0.1-2\,keV with a time binning of 1000\,s and $1\sigma$ Poissonian error bars.}
   \label{individual_lcs_sec}%
   \end{figure}

We detect the stellar companion HD~189733B in X-rays for the first time. It displays a mean count rate of $\approx3$\,cts\,ks$^{-1}$, with considerable variability by up to a factor of 5. The fourth light curve (Fig.~\ref{individual_lcs_sec}) displays a prominent flare-like variation which lasts for ca.\ 6\,ks. Variability on shorter time scales is also implied by the other light curves, but individual small flares cannot be characterized properly due to the low count rate.

The spectra of HD~189733B are well described by a thermal plasma with two temperature components, with a dominant temperature contribution at ca. 3.5\,MK (see Table~\ref{spec_AB}). The collected spectral counts are not sufficient to constrain elemental abundances in the corona, and therefore we fixed the abundances at solar values. The spectral fit yields an X-ray luminosity of $\log L_X = 26.67$\,\ergs \ in an energy range of $0.25-2$\,keV. This is compatible with the upper limit of $\log L_X \leq 27$\,\ergs \ derived from {\it XMM-Newton} observations by \cite{Pillitteri2010}. In the context of late-type stars in the solar neighborhood, HD~189733B is slightly less X-ray luminous than the median value of $\log L_X = 26.86$\,\ergs \ found for early M dwarfs \citep{Schmitt1995}. To calculate the bolometric luminosity of HD~189733B, we use the relations from \cite{Worthey2011}: the measured $H-K$ color of 0.222 and a spectral type between M3.5V and M5V \citep{Bakos2006} translates to a bolomeric correction of ${\rm BC_V} = -2.05$ to a V magnitude of 14.02, yielding $L_{bol} = 1.8\times10^{31}$\,\ergs. The X-ray fraction of the total bolometric flux is therefore $R_X = L_X / L_{bol} = 2.5\times10^{-5}$, marking HD~189733B as a moderately active M star.

\subsection{The background source}

We detect X-ray emission from a source at the position ${\rm R.A.} = 300.1813$, ${\rm decl.} = 22.7068$, previously reported by \cite{Pillitteri2010}. The projected distance to HD~189733A is $11.7^{\prime\prime}$, the position angle is ca.\ 190$^\circ$. The infrared 2MASS data shows a filter glint from HD~189733A at this position, so that the infrared magnitudes of the source are unknown; however, other surveys find two faint objects at the approximate position, see \cite{Bakos2006} and their Figure~1. 

\begin{figure}[h!]
   \centering
   \includegraphics[width=0.5\textwidth]{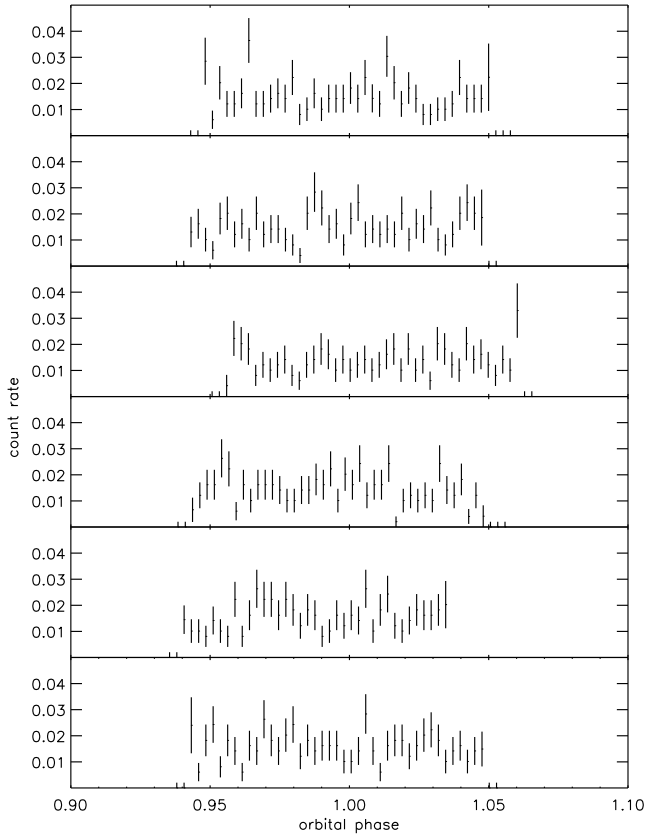}
   \caption{Individual X-ray light curves of the third source, taken with {\it Chandra} ACIS-S. All light curves are extracted from an energy range of 0.1-10\,keV with a time binning of 500\,s and $1\sigma$ Poissonian error bars.}
   \label{individual_lcs_unk}%
   \end{figure}

The X-ray emission from this source is at a mean level of 0.015\,cts\,s$^{-1}$ and varies over time by factors of up to 2.5, see Fig.~\ref{individual_lcs_unk}; the fifth light curve displays a variability feature that looks similar to a stellar flare. The X-ray spectra from the source show no significant flux at energies below 800~eV, as shown in Fig.~\ref{spectra}, but the flux at high energies ($> 2$\,keV) is larger than for HD~189733A and HD~189733B. A spectral feature is present at energies near 1.9\,keV; additional smaller features may be present near 1.4\,keV and 4\,keV. The feature at 1.9\,keV matches with emission lines of Si\,{\sc xiii/xiv} between 1839-2005\,eV; the 1.4\,keV feature coincides with several strong Mg\,{\sc xi/xii} lines between  1352-1472\,eV, while the 4\,keV feature falls into a spectral region where some weaker Ca\,{\sc xix/xx} lines are present.

\begin{table}[t!]
\begin{tabular}{l l l}
\hline \hline
model parameters				& abs.\ thermal plasma		& abs.\ power law	\\ \hline
$n_H$						& $(4.5\pm0.6)\times 10^{21}$	& $(5.8\pm 0.7)\times 10^{21}$	\\
$T$ (keV)					& $6.0\pm 0.9$			&	-	\\
photon index					&	-			& $1.9\pm 0.1$				\\ \vspace{0.2cm}
norm						& $(1.1\pm0.06)\times 10^{-4}$	& $(3.9\pm0.5)\times 10^{-5}$	\\ \vspace{0.2cm}
$\chi^2_{red}$ (d.o.f.)				& 1.12 (99)			& 1.11 (99)	\\
$F_X$ (0.25-10 keV)				& $1.44\times 10^{-13}$		& $1.47\times 10^{-13}$		\\
$\log L_X^{abs}$ (0.8-10 keV)			& $31.2$			& $31.2$		\\ 
$\log L_X^{unabs}$				& $31.4$			& $31.4$		\\ \hline
\end{tabular}
\caption{Spectral fits for the X-ray emission of the third source. An absorbed thermal plasma model and an absorbed power law both yield good fits. The absorbed and unabsorbed X-ray luminosities are calculated using a distance estimate of $1000$\,pc.}
% assume 1 cm-3 as the mean ISM density -> 1600 pc
\label{spec_unk}
\end{table}

The missing flux at soft X-ray energies and the spectral feature at 1.9\,keV suggest that the object is a distant source which can be described as a thermal plasma with interstellar absorption. Indeed, the corresponding model yields a good fit to the data as shown in Table~\ref{spec_unk}. The modelled temperature of ca.\ 70\,MK is most likely biased towards high values, because plasma of lower temperatures is not seen due to the interstellar absorption and can thus not adequately be fitted. The peak formation temperatures of the Si\,{\sc xiii/xiv} lines are at a range of 10-17\,MK; we interpret this as an indication that the average plasma temperature is significantly lower than 70\,MK. We assume from the fitted hydrogen absorption with a column density of $n_H = 4.5\times 10^{21}$\,cm$^{-2}$ that the source is located at a distance of roughly 1000\,pc from the Sun, consistent with \cite{Pillitteri2010}. With this distance, the measured flux converts to an intrinsic (unabsorbed) X-ray luminosity of $\log L_X = 31.4$\,\ergs. 

So what is the nature of this source? Its position in galactic coordinates is $l = 60.9623$, $b = -03.9227$, meaning that it lies in the galactic plane in a viewing direction 60$^\circ$ off the galactic center, possibly on the Orion-Cygnus arm. Together with the rough distance estimate from hydrogen absorption, the emission feature in the spectra and the flare-shaped variation in the light curves, we consider it likely that the source is an RS~CVn or Algol system. Its X-ray luminosity is at the high end, but still in the range of observed values for such systems \citep{Walter1980, White1983}.

For completeness, we note that a model with an absorbed power law formally yields an equally good fit, but does not explain the observed feature at 1.9\,keV. Adopting the power law model, we find similar values for hydrogen absorption and X-ray luminosity (see Table~\ref{spec_unk}).

\subsection{The phase-folded X-ray light curve of HD 189733A}\label{phasefold}

We now turn to a detailed analysis of the phase-folded and added up X-ray transit light curve of the Hot Jupiter HD~189733Ab.

\subsection{Testing for autocorrelation and red noise}

As a first step, we test if the scatter in the observed individual and added-up light curves is sufficiently well described by white noise. If the sequence of measured count rates is not randomly distributed on time scales shorter than the transit, but is correlated in some way, this would require a more detailed treatment of the noise level. We have therefore tested for autocorrelation in the individual light curves, as well as for deviations from white noise in the added-up light curve, both with negative results. This analysis is presented in detail in the Appendix.

\subsubsection{A transit detection in X-rays}

\begin{table*}
\noindent\makebox[\textwidth]{
\begin{tabular}{l c c c c}
\hline \hline
						&(a) all 7 data 		& (b) 6 data sets,  		& (c) 6 data sets, 		& (d) 5 non-flaring  	\\ 
						& sets				& excluding flaring 		& excluding flaring 		& data sets		\\
						&				& {\em Chandra} observation	& {\it XMM-Newton} observation	&	  		\\ \hline
raw count rate in transit (cts\,$s^{-1}$)	& $\mathbf{0.4696\pm0.0106}$	& $0.4122\pm0.0099$		& $0.3249\pm0.0088$		& $\mathbf{0.2675\pm0.0080}$	\\
raw count rate out of transit (cts\,$s^{-1}$)	& $\mathbf{0.4996\pm0.0075}$	& $0.4467\pm0.0071$		& $0.3508\pm0.0063$		& $\mathbf{0.2979\pm0.0058}$	\\
$\chi^2$ of constant model			& $\mathbf{5.41}$		& $8.08$			& $5.79$			& $\mathbf{9.58}$		\\
constant model rejected at confidence		& $\mathbf{98.00\%}$		& $99.55\%$			& $98.39\%$			& $\mathbf{99.80\%}$		\\
\end{tabular}
}
\caption{Added-up raw count rate (i.e. without pre-normalization) in and out of transit, for the four combinations of data sets discussed in section~\ref{lcselection}, together with the confidence level at which a constant model is rejected.}
\label{transitdetection}
\end{table*}

As a second step, we investigate if the transit is detected in the soft X-ray light curves by testing if the null hypothesis of a constant X-ray count rate in and out of transit can be rejected. To use as few assumptions as possible, we use only orbital phases during which all of the observations provide data for this; specifically, we use orbital phases between $0.9560$ and $1.0345$. This makes it possible to compare the raw (i.e.\ unnormalized) number of detected soft X-ray photons in and out of transit. For the {\em XMM-Newton} observations we use the net number of source photons, i.e.\ the source signal corrected for the low-level background signal. We define orbital phases between second and third contact as ''in transit'' and phases before first or after fourth contact as ''out of transit''. We then add up the detected X-ray photons of all seven light curves in and out of transit, divide by the duration of the in and out of transit times, and compute the $\chi^2$ goodness of fit to a constant model with the overall average photon rate. We list the in and out of transit count rates, the $\chi^2$ value, and the confidence at which the null hypothesis of a constant count rate can be rejected in Table~\ref{transitdetection}. We find that for all seven datasets combined, the null hypothesis can be rejected at 98.00\% confidence. For other combinations of five or six datasets (which are motivated below in section~\ref{lcselection}), we find confidence levels between 98.39\% and 99.80\%. Having shown that we have indeed detected the transit of the exoplanet in X-rays with high confidence, we proceed by deriving the transit depth.

\subsubsection{Deriving the transit depth}

We have one observation that starts relatively late (Chandra light curve 3) and one that ends early (Chandra light curve 5). To increase our accuracy for a transit depth measurement, we therefore pre-normalized the individual transit light curves to an out-of-transit level of unity and use a larger orbital phase range between $0.9479$ and $1.0470$, as described in section~\ref{phasedlc}.

We used three different statistical methods to determine the depth of the X-ray transit in the phase-folded light curve; they are explained in the following and their results are listed in Table~\ref{transittable}.

(1) The mean normalized X-ray count rates in and out of transit were compared directly. We have defined data from orbital phases between second and third contact as ''in transit'' and data from phases before first or after fourth contact as ''out of transit''. Since the total number of X-ray photons is large, differences between Poissonian and Gaussian errors are negligible, and we have used Gaussian error propagation to determine the error on the transit depth as $\sigma_{depth} = \sqrt{ \sigma^2_{in} + \sigma^2_{out} }$.

\begin{figure*}[t!]
   \centering
   \subfigure[All seven data sets]{\includegraphics[width=0.45\textwidth]{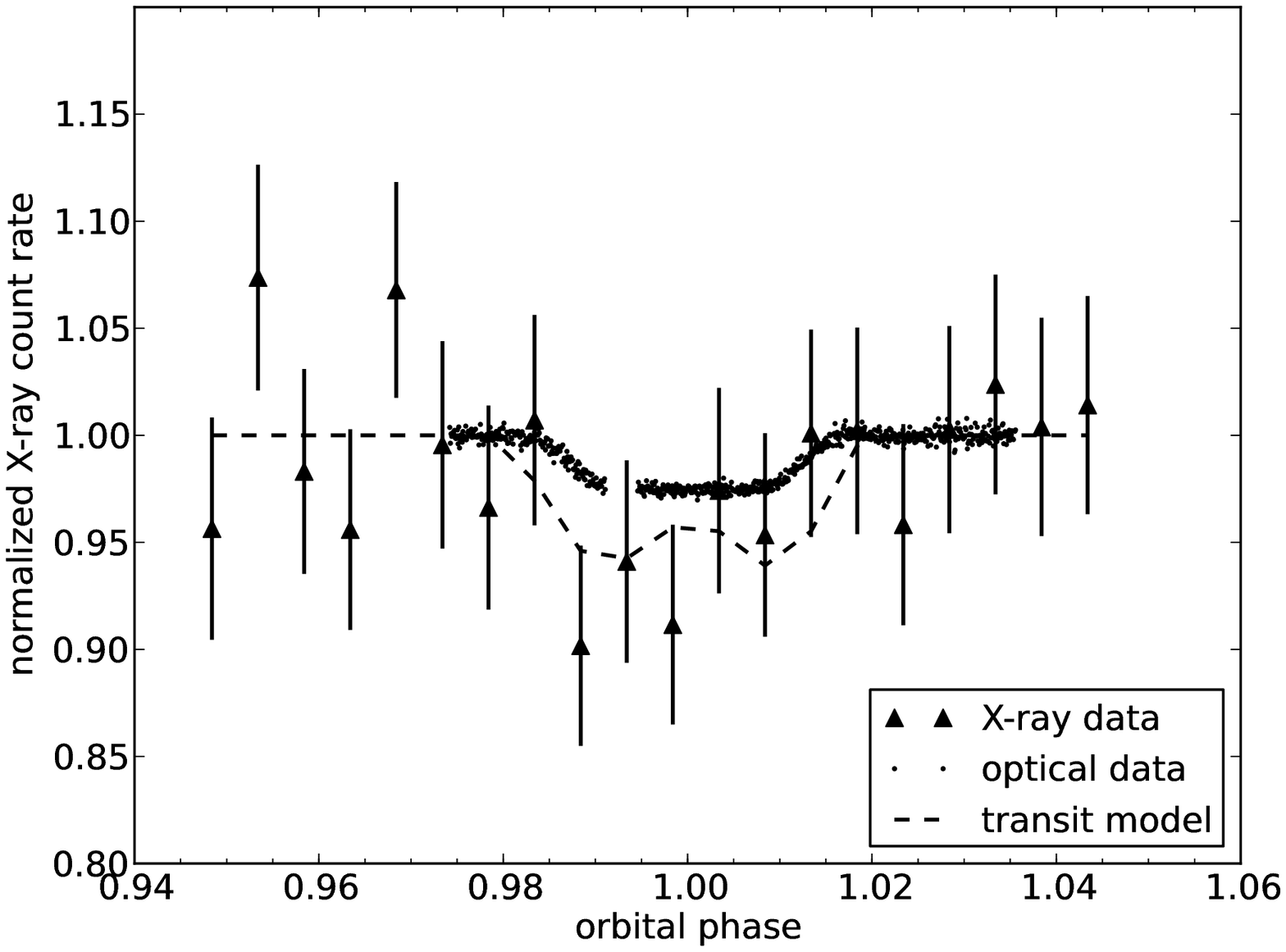} }
   \hspace{0.5cm}
   \subfigure[Six data sets, excluding the potentially flaring {\it Chandra} observation]{\includegraphics[width=0.45\textwidth]{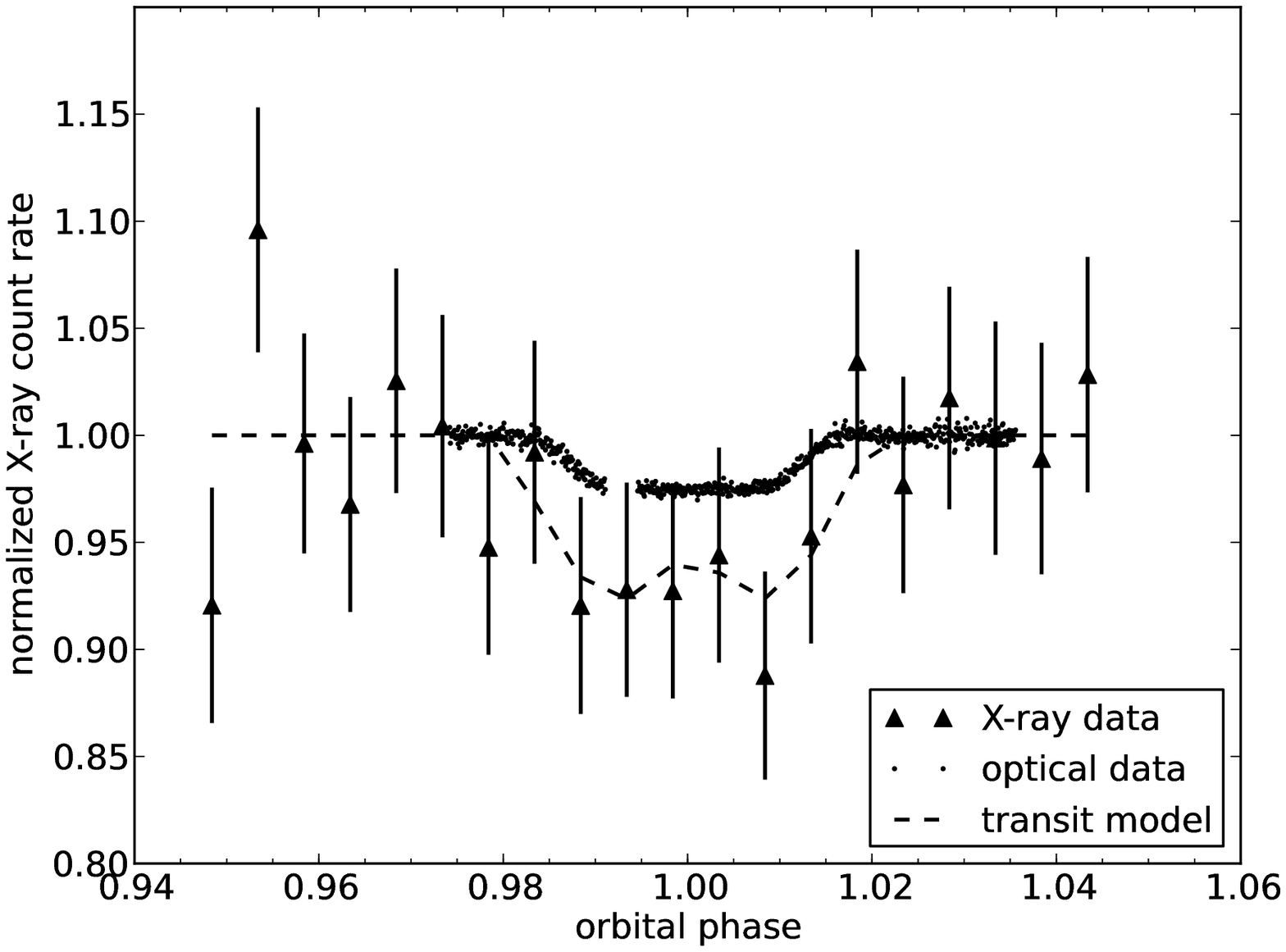} }
   \subfigure[Six data sets, excluding the potentially flaring {\it XMM-Newton} observation]{\includegraphics[width=0.45\textwidth]{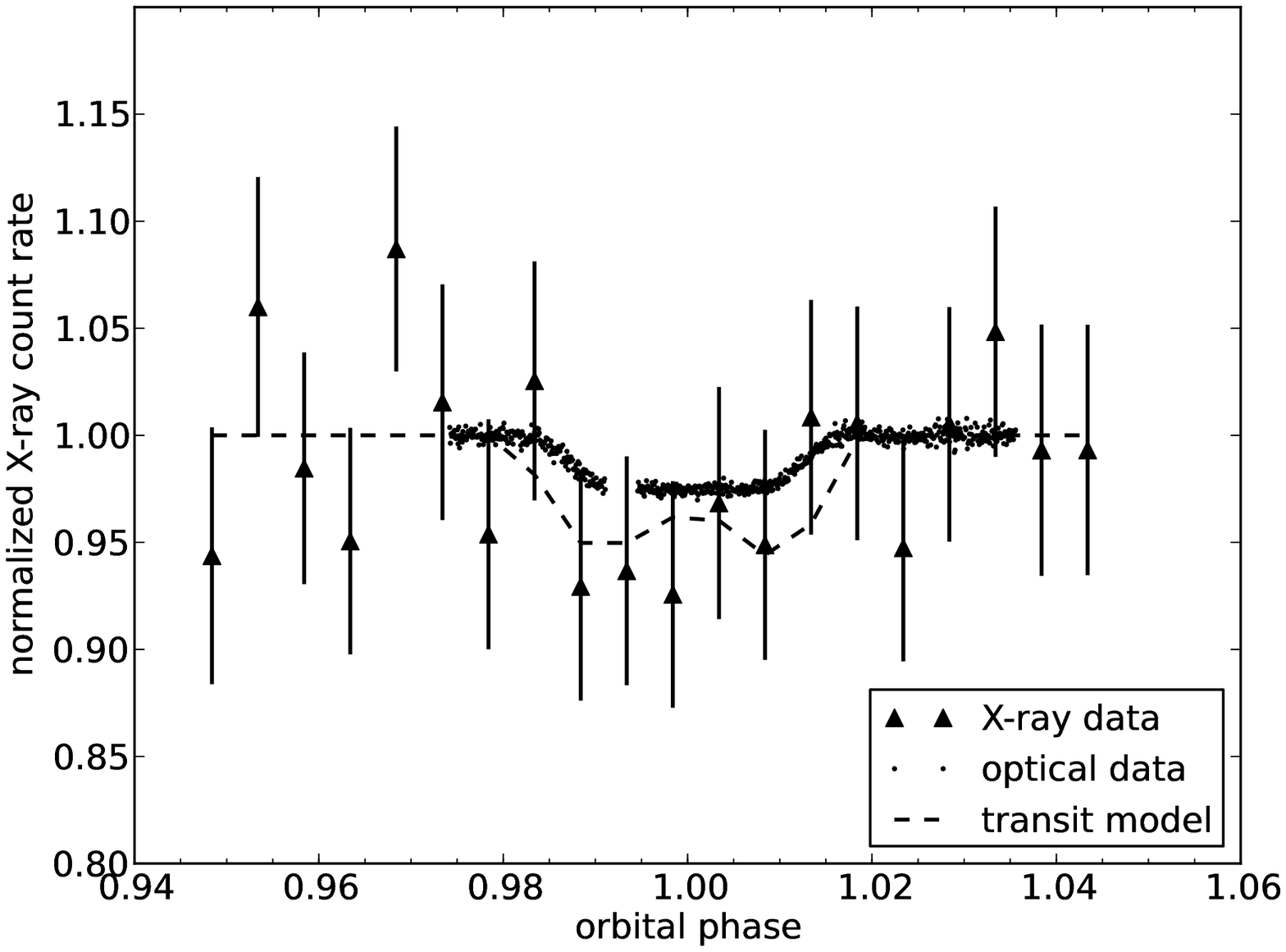} }
   \hspace{0.5cm}
   \subfigure[Five quiescent data sets, excluding both potentially flaring observations]{\includegraphics[width=0.45\textwidth]{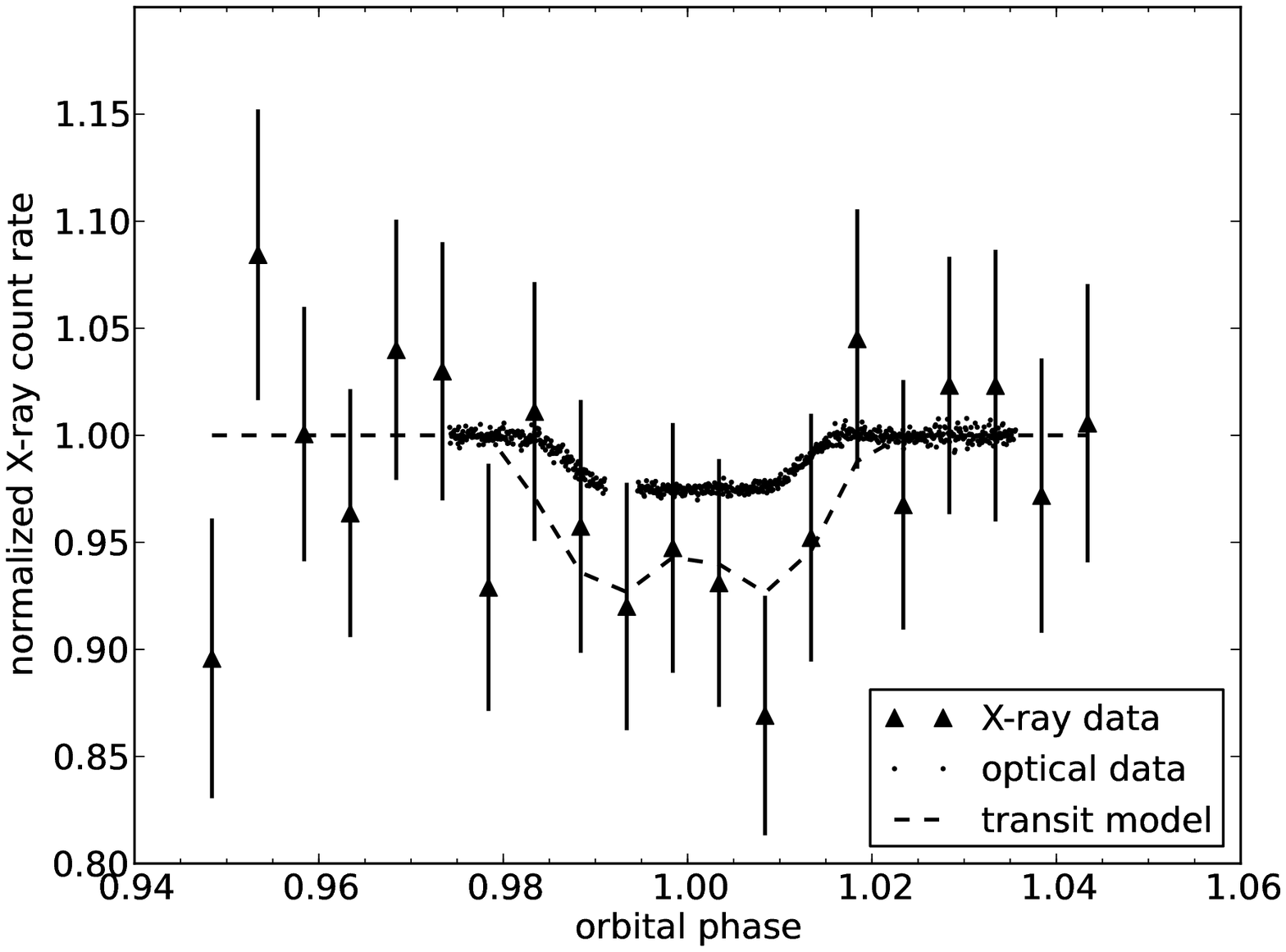} }

   \caption{The X-ray transit in comparison with optical transit data from Winn et al.~(2007); vertical bars denote $1\sigma$ error bars of the X-ray data, dashed lines show the best fit to a limb-brightened transit model from \cite{Schlawin2010}. The X-ray data is rebinned to phase bins of 0.005. The individual figures show different data combinations.}
   \label{XrayLC}%
   \end{figure*}

(2) We have fitted the data to an analytical transit model as used for optical, limb-darkened transits \citep{MandelAgol2002}. This will not be entirely correct, because coronae are limb-brightened, not limb-darkened like photospheres (see method 3). However, we give the fitting results to photospheric transit profiles for completeness. Specifically, we fixed the planetary orbital period, semimajor axis, mid-transit time, and inclination to the values given in the Extrasolar Planets Encyclopedia, used a quadratic limb darkening law with coefficients $u1 = 0.32$ and $u2 = 0.27$ as used for the optical light curve \citep{Winn2007} depicted in Fig.~\ref{XrayLC}(a), and performed a Markov-Chain Monte Carlo fit with the ratio of the planetary and stellar radius $R^{X-ray}_P/R_\ast$ as a free parameter.

(3) The stellar corona is optically thin, so that it shows limb {\em brightening} instead of the limb darkening observed in the photosphere. We have taken this into account by fitting the data with a limb-brightened model \citep{Schlawin2010} under the assumption that the corona is not significantly extended beyond the photospheric radius. This is a valid assumption: It is known that individual flaring loops in active stars may extend quite far over the photospheric radius, but the average corona of active stars apparently is not much extended. This was directly observed in the case of the binary $\alpha$~CrB, where the X-ray dark A star eclipses the active, X-ray bright G star without significant timing differences from the optical transit profile \citep{KuersterSchmitt1996}. We again fixed the planetary orbital period, semimajor axis, mid-transit time, and inclination, and fitted the ratio of the planetary and stellar radius. The resulting transit profile is shown as a dashed line in Fig.~\ref{XrayLC}(a).

\begin{table*}
\noindent\makebox[\textwidth]{
\begin{tabular}{l c c c c}
\hline \hline
				&(a) all 7 data 		& (b) 6 data sets, excluding 	& (c) 6 data sets, excluding 	& (d) 5 non-flaring  	\\ 
				& sets				& flaring {\em Chandra} observation& flaring {\it XMM-Newton} observation& data sets		\\ \hline
transit depth derived from: 	& 				& 				& 				&			\\
(1) direct comparison		& $\mathbf{0.059\pm0.026}$	& $0.080\pm0.028$		& $0.057\pm0.030$		& $\mathbf{0.081\pm0.032}$	\\
(2) limb-darkened model		& $0.057\pm0.028$		& $0.082\pm0.029$		& $0.048\pm0.030$		& $0.077\pm0.034$	\\ 
(3) limb-brightened model	& $\mathbf{0.050\pm0.025}$	& $0.073\pm0.025$		& $0.040\pm0.027$		& $\mathbf{0.067\pm0.030}$	\\ 
\end{tabular}
}
\caption{X-ray transit depths, derived from the added-up, prenormalized X-ray light curves. The results are given for the four different combinations of data sets discussed in section~\ref{lcselection}. The transit depths are given with $1\sigma$ errors (statistical).}
\label{transittable}
\end{table*}

\subsubsection{Light curve selection}\label{lcselection}

The stellar corona is much less homogeneous than the stellar photosphere. X-ray emitting coronal loops are temporally variable, and solar observations show that its corona often displays several large, X-ray bright patches, while other parts are dimmer in X-rays. This is why it is unlikely to identify a planetary transit signal in a single X-ray light curve, even if the signal to noise is very high: It is possible that, given a specific configuration of X-ray bright patches of the corona, the planet happens to transit a mostly X-ray dim part. On the other hand, the planet may occult a large X-ray bright patch in a different coronal configuration, causing an untypically deep transit signal. One therefore needs to average over as many transits as possible to determine the transit depth with respect to the typical, average X-ray surface brightness of a corona. 

\begin{figure}[t!]
   \centering
   \includegraphics[width=0.45\textwidth]{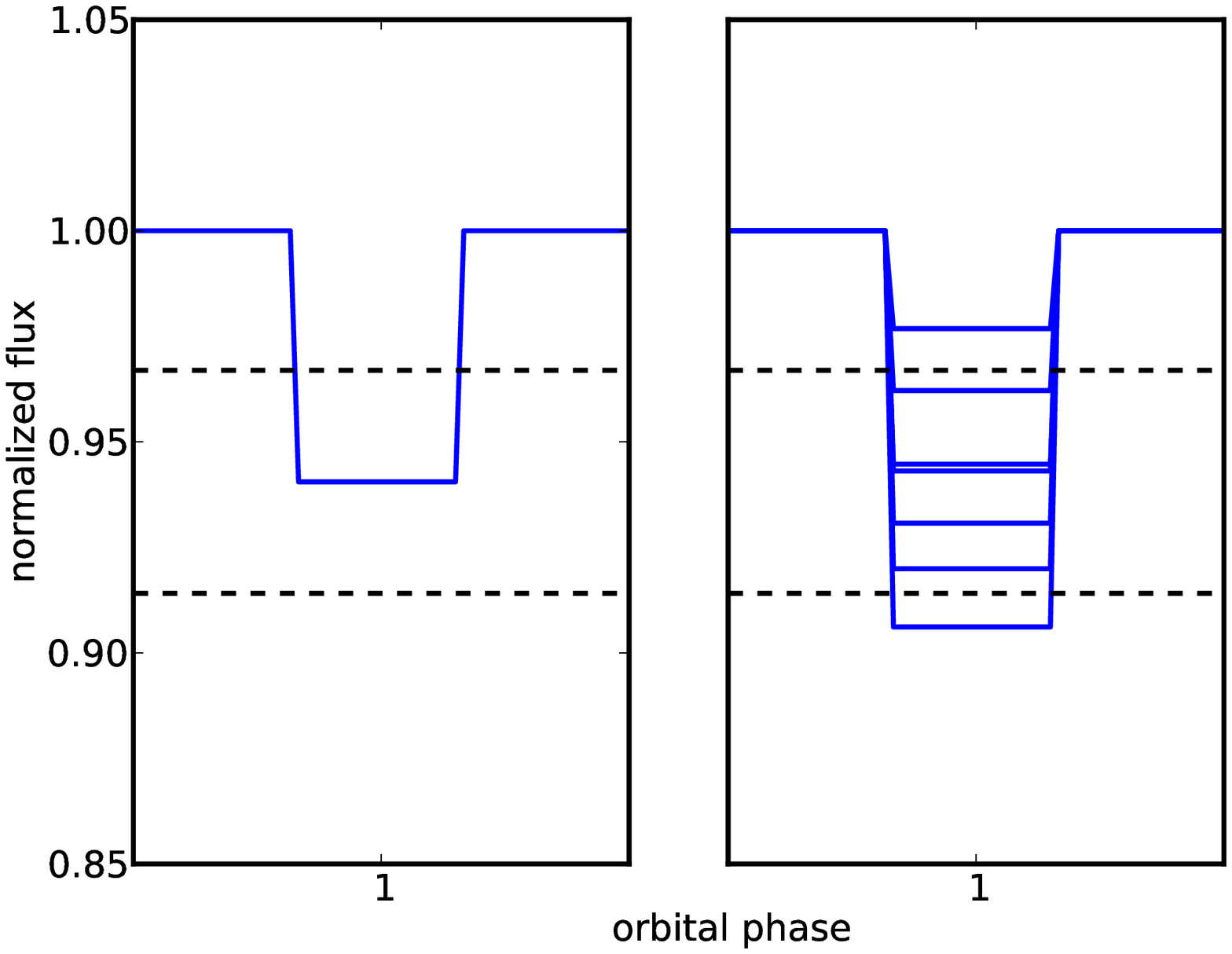}
   \caption{Symbolical transit light curves depicting the X-ray transit depth derived from all seven data sets (left), and from the seven possible combinations which each include six data sets (right). For better visibility, any limb effects are ignored.}
   \label{depth_scatter}
   \end{figure}

To show the importance of the coronal averaging, we determined the observed transit depths (derived from the direct count rate comparison) for different combinations of the X-ray light curves. With a total of seven transit light curves, there is one combination which includes all light curves, and seven combinations which include six of the light curves. Given the relatively low overall number of X-ray photons, we do not use combinations of fewer light curves because the errors become quite large. We show the result in Fig.~\ref{depth_scatter}. For better visibility, we symbolically show the mean transit depth without depicting any limb effects. The transit depths derived from only six X-ray light curves show some scatter; specifically, the depths range from 2.3\% to 9.4\%. Two out of seven (i.e.\ 29\%) depth measurements lie outside the nominal error derived from all seven light curves ($5.9\%\pm 2.6\%$). The differences we find by excluding single light curves reinforces our notion that we are likely dealing with a patchy stellar corona, and individual X-ray transit light curve may show strongly differing transit depths. It would therefore be desirable to average over a very large number of X-ray transits. However, we are limited to an analysis of the available seven light curves, in which the inhomogeneities of the stellar corona may not have been fully avaraged out.

There are a few combinations of six light curves that are worth investigating further. One is the combinations of all six {\it Chandra} light curves, excluding the {\it XMM-Newton} light curve. {\it XMM-Newton} and {\it Chandra} are very well cross-calibrated, so one does not expect systematic differences here. And indeed, the transit depth derived from the {\it Chandra} light curves alone is $5.7\%$.

Another combination that we want to look at in more detail is all light curves except the second {\it Chandra} light curve. From visual inspection of the light curves, it seems that there may be a small flare in the second {\it Chandra} light curve near phase $1.01-1.02$. It is rather difficult to identify a flare unambiguously in light curves of this signal to noise level. Further, \cite{Pillitteri2011} have observed that small flares of HD~189733A are often {\it not} accompanied by a significant change in hardness ratio (which is often seen in larger flares on cool stars). Rather, they observed an almost simultaneous rise in both the hard and the soft X-ray band, with only a slightly earlier increase in hard X-rays. To test for such a behavior in our light curves, we calculate the Stetson index for 4~ks box car chunks for each light curve. The Stetson index measures correlated changes in two light curve bands, with positive values indicating correlated changes, negative values indicating anti-correlated changes, and values close to $0$ indicating random changes; see \cite{WelchStetson1993}. \cite{Carpenter2001} showed the 99\% threshold for (anti)correlated variability is achieved near $|S|=0.55$. A more conservative value of $|S|>1$ is used to identify stronger variability \citep{Rice2012}. 

   \begin{figure}[t!]
   \centering
   \includegraphics[width=0.45\textwidth]{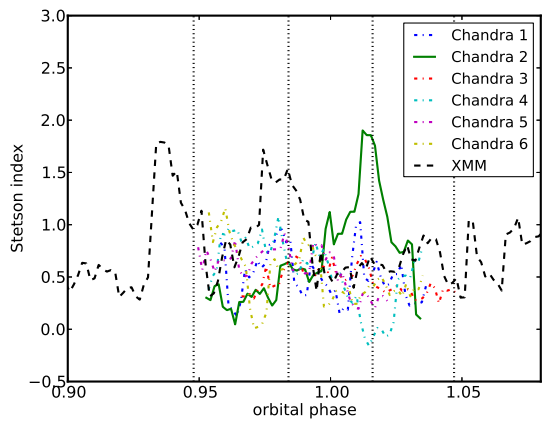}
   \caption{Stetson indices of the individual light curves, using a soft (0.2-0.6\,keV) and hard (0.6-2.0\,keV) X-ray band for box cars of 4\,ks duration. The {\it XMM-Newton} light curve and the second {\it Chandra} light curve show the strongest correlated changes in the two bands. The vertical lines again depict the data stretches used for the combined light curve (outer lines) and the first and fourth contact of the optical transit (inner lines).}
   \label{stetson}
   \end{figure}

We show the result of this analysis in Fig.~\ref{stetson}. All light curves display al least mildly correlated changes in the soft and hard X-ray band, and two light curves display strongly significant correlated variability during part of the observations. Those light curves are from the {\it XMM-Newton} observations, shown as dashed black line, and the second {\it Chandra} observation, shown as solid green line. For the {\it Chandra} observation, the phase of significant correlated variability is the same as identified by visual inspection, namely around $1.01-1.02$. For the {\it XMM-Newton} light curve, one of the correlated phases coincides with a flare-shaped part of the light curve around phase $0.94$ (outside the phase interval considered for the added-up light curve). No flare-like feature is seen for the other part of that light curve with correlated changes (phase $0.97-0.98$), but a slight downward slope with small variability features is present. We interpret these findings as consistent with small flares occurring during the second {\it Chandra} observation and the first half of the {\it XMM-Newton} observation.  

We therefore identify four light curve combinations which are of interest:  (a) all data sets (total: six observations from {\it Chandra} and one from {\it XMM-Newton}),  (b) six data sets, excluding the {\it Chandra} observation with the assumed flare (total: five observations from {\it Chandra} and one from {\it XMM-Newton}),  (c) only the six {\it Chandra} data sets, excluding the potentially flaring {\it XMM-Newton} observation, (d) only the five quiescent {\it Chandra} data sets, excluding the potentially flaring {\it Chandra} and {\it XMM-Newton} observations. This last combination will have rather large errors because two out of seven observations are ignored, but the flare analysis conducted above demonstrates the necessity to investigate this combination as well.

We list the transit depths derived for the three additional light curve combinations in Table~\ref{transittable}, and show the limb-brightened fits to the data in Fig.~\ref{XrayLC}(b)-(d).

These results show that excluding the {\it XMM-Newton} data set has little consequence for the derived transit depth, but excluding the possibly flare-contaminated second {\it Chandra} light curve changes the depth by about 35\%. This is still within the $1\sigma$ error range derived from Poissonian counting statistics. Given our flare analysis of the individual light curves above, we think that the true X-ray transit depth is closer to a value of $8\%$ derived from the non-flaring observations than to $5.9\%$ derived from all observations. We have therefore bold-faced two columns in Table~\ref{transittable}: column one representing the transit depth derived from all data sets, without selecting for stellar quiescence, and column four, representing the transit depth derived from only the quiescent data sets. We would like to remind the reader that our excess transit depths may be influenced by inhomogeneities of the stellar corona. We will discuss the possibility of {\em latitudinal} inhomogeneities in section~\ref{latitudes}, which could be induced by activity belts. However, we have no means of testing for {\em longitudinal} inhomogeneities, i.e.\ testing whether the planet has occulted a typical number of coronal active regions during these specific seven observations. Additional X-ray observations are therefore necessary to confirm that the available X-ray data indeed sample a representative average of the stellar corona.

\section{Discussion}

\subsection{What causes the deep X-ray transit?}

There are two possible explanations for the observed transit depth: in scenario A, the planet's X-ray and optical radii are more or less the same, but the planet occults parts of the stellar corona considerably brighter than the disk-averaged X-ray emission; in scenario B, the effective X-ray radius of the planet is significantly larger (by $>50\%$) than its optical radius, and the occulted parts of the stellar corona have the same or a smaller X-ray brightness than the disk average. 

\subsubsection{Scenario A: An inhomogeneous corona}\label{latitudes}

This scenario requires active regions to be present in the transit path. In cool stars, such active regions are bright in X-rays and dark in the photosphere. We can rule out at once that the planet occults {\em the same} active region in the observed transits, because the six observed {\it Chandra} transits were spread out over 18~days, while the stellar rotation period is about 12~days \citep{Henry2008}. However, similar to the Sun, HD 189733 might possess an active region belt making the transit path brighter than average in X-rays. We do not have simultaneous optical coverage to our X-ray data, but we can use optical data from earlier epochs to derive a general picture of the active region distribution on the host star. The overall spot coverage can be inferred from long-term rotational light curve modulation, if the contrast between the star spots and the unspotted photosphere is known. White-light observations with the MOST space telescope ($\sim$ 4000-7000~\AA) showed that HD~189733 displays an out-of-transit rotational moduation with an amplitude of the order of $\pm 1.5\%$ \citep{Miller-Ricci2008}. Spot temperatures of 4250\,K compared to a mean photospheric temperature of 5000\,K have been inferred from two sets of HST observations at optical and near-infrared wavelengths at $\sim$ 2900-5700~\AA\ and 5500-10500~\AA \citep{Pont2007, Sing2011}. Assuming that the observed spot temperatures are typical for the spots on this star, these spots are 60\% darker than the photosphere in the MOST white light band, and 70\% (50\%) darker in the optical (infrared) HST band. Therefore the minimal areal spot coverage of the stellar disk needed to induce the observed rotational modulation in MOST is $\approx 5\%$. 

The important question then is if the transit path, which has an area of 13\% of the stellar disk, contains a larger spot fraction then the average photosphere. The spot coverage in the transit path can be inferred from signatures of occulted star spots during transits. Several spot occultations have been observed for HD~189733 \citep{Pont2007, Sing2011}; the spot sizes can be estimated from the amplitude and duration of the light curve features and the contrasts in the respective wavelength bands. \cite{Pont2007} report two starspot occultations during two different transits, with inferred spot sizes of 80\,000\,km\,$\times$\,12\,000\,km for the first spot and a circular shape with a radius of 9000\,km for the second spot. This translates to a transit spot coverage fraction of 0.8\% for the first transit and 0.2\% for the second transit. \cite{Miller-Ricci2008} report a nondetection of spot signatures in MOST observations of HD~189733Ab transits; their upper limit to spot sizes is equivalent to circles with a radius of $0.2\,R_{Jup}$, corresponding to a transit path spot coverage of $\lesssim 0.5\%$. An occultation of a large spot was observed by \cite{Sing2011}. Those authors do not give a spot size estimate, but it can be derived to be ca.\ 3.3\%.\footnote{The spots are 70\% darker than the photosphere in the used band; the spot causes an upward bump of ca.\ 0.003 compared to the unspotted part of the transit light curve, with the out-of-transit level being unity. During transit, the bump is caused by the spot ''appearing'' and ''disappearing'' on the stellar disk; thus, the spot size is roughly $0.003/0.7 \times \pi R^2_{\ast} \approx 6\times10^{9}$\,km$^2$.} Combining the information from all those observations, we can conclude that the transit path of HD~189733b does not have a significantly larger spot coverage fraction than the rest of the star, making scenario A unlikely as an explanation for the observed transit depth in X-rays.

\subsubsection{Scenario B: An extended planetary atmosphere}

This scenario requires that HD~189733b possesses extended atmosphere layers which are optically thin at visual wavelengths, but optically thick to soft X-rays. The stellar X-ray spectrum has a median photon energy of about 800~eV. At those energies, the main X-ray absorption process is photoelectric absorption \citep{Morrison1983} by metals such as carbon, nitrogen, oxygen, and iron, despite their small number densities of $\sim 10^{-4}-10^{-5}$ compared to hydrogen in a solar-composition gas. Hydrogen contributes very little to the total X-ray opacity because the typical energy of the stellar X-ray photons is much larger than the hydrogen ionization energy of 13.6~eV. For a solar-composition plasma, column number densities of $2\times10^{21}$\,cm$^{-2}$ are needed to render the plasma opaque (optical depth $\tau = 1$) to soft X-rays.

We will now perform some oder-of-magnitude estimates for the properties of the planetary atmosphere; we will use an X-ray transit depth of 8\% for this, i.e.\ the value derived from quiescent datasets, while noting that smaller transit depths are also compatible with the data. An X-ray transit depth of ca.\ 8\% and thus an effective planetary X-ray radius of $R_X \approx 1.75\times R_P \approx 1.4\times10^{10}$\,cm requires atmospheric densities of $\sim7\times 10^{10}$\,cm$^{-3}$ at $R_X$. If the metallicity of the planetary atmosphere is higher - either overall or selectively at higher atmosphere layers - the required densities for X-ray opaqueness decrease roughly linearly.

The overall metal content of HD~189733b is unknown, even if some individual molecules have been detected in its lower atmosphere \citep{Tinetti2007}. In the solar system, Jupiter and Saturn are metal-rich compared to the Sun; several models for those two planets agree that the metallicity is at least three times higher than in the Sun \citep{FortneyNettelmann2010}. For the Hot Jupiter HD~209458b a substellar abundance of sodium \citep{Charbonneau2002} has been found, but other observations \citep{Vidal-Madjar2004, Linsky2010} are compatible with solar abundances. However, it is important here that the X-ray flux at the orbit of HD~209458b \citep{Sanz-Forcada2011} is lower compared to HD~189733b by two orders of magnitude ($<40$\,erg\,cm$^{-2}$\,s$^{-1}$ vs.\ $5700$\,erg\,cm$^{-2}$\,s$^{-1}$), so that the upper atmospheres of these planets may be very different. For HD~189733b, its host star has solar metallicities \citep{Torres2008}, while observations of sodium in the planetary atmosphere \citep{Huitson2012} are compatible with various abundances depending on the details of the atmosphere.  Atmospheric stratification processes might cause deviations from the bulk metallicity in the upper atmosphere of HD~189733b; this is observed for the Earth, which displays an enrichment of O$^{+}$ in the higher ionosphere, because the planetary magnetic field increases the scale height of ions compared to neutral species \citep{Cohen2012ambipolar}. 

Determining the atmospheric structure theoretically would require a self-consistent model of the planetary atmosphere which is beyond the scope of this article. However, we can test which temperatures are necessary to provide sufficient densities at $R_X$. The density, and therefore temperature, needed in the outer planetary atmosphere to be X-ray opaque depends on the assumed metallicity at $R_X$. In the following, we will assume a moderate metal enrichment in the planetary atmosphere of ten times solar metallicity,i.e.\ three times the bulk metallicity of Jupiter. This yields a required density of $7\times 10^{9}$\,cm$^{-3}$ at the X-ray absorbing radius of $R_X \approx 1.75\times R_P$. The required high-altitude density can be compared to observed densities in the lower atmosphere, assuming a hydrostatic density profile. Models show that this assumption is valid \citep{Murray-Clay2009} with very small deviations below the sonic point at ca.\ 2 to 4 $R_P$.

The observational determination of densities in exoplanetary atmospheres is challenging. HST observations suggest that the planetary radius slightly increases with decreasing wavelength between 300 and 1{,}000~nm \citep{Sing2011, Pont2008}, which is interpreted as a signature for Rayleigh scattering in the planetary atmosphere. If molecular hydrogen is the dominant scattering species, a reference pressure on the order of $\approx150$~mbar results at a temperature of $\sim 1{,}300$~K; if, however, the scattering is dominated by dust grains, the reference pressure might be considerably smaller (on the order of $0.1$~mbar \citealt{Lecavelier2008Rayleigh}). Recently, a sudden temperature increase to $2{,}800$~K was observed in HD~189733b's lower atmosphere \citep{Huitson2012}. The corresponding atmosphere layer has a pressure of $6.5-0.02$~mbar (particle number densities $\sim 2\times 10^{16} -  5\times 10^{13}$~cm$^{-3}$) for the Rayleigh scattering model, and $4\times 10^{-3} - 10^{-5}$~mbar (particle number densities $\sim 10^{13} -  3\times 10^{10}$~cm$^{-3}$) if dust scattering dominates. Temperatures at higher altitudes have been modelled to be on the order of $10{,}000-30{,}000$~K \citep{Yelle2004, Penz2008}.

We can determine which temperatures are required to provide sufficient densities at $R_X$, assuming a hydrostatic atmosphere below $R_X$ and using the high and low reference pressures given above. For dust scattering, the pressure is too low to provide the necessary densities of $7\times 10^{9}$~cm$^{-3}$ at $R_X$ with reasonable temperatures. However, for the higher pressure of the Rayleigh scattering case, a scale height of $H = 4{,}200 - 6{,}800$~km provides sufficient density at $R_X$ to match our observations. The required temperature can be extracted from the scale height $H = \frac{kT}{\mu g}$, with $g$ being the planetary gravity and $\mu$ the mean particle weight, which we choose to be $1.3$ times the mass of the hydrogen atom because molecular hydrogen dissociates at high temperatures. This estimate yields temperatures of $\sim 14{,}000 - 23{,}000$~K which need to be present in the upper atmosphere, consistent with high-altitude temperatures found by theoretical models \citep{Yelle2004, Penz2008}. 

At temperatures higher than $10,000$~K hydrogen will be mostly ionized. In a low-density ($<10^{14}$~cm$^{-3}$) plasma, which is true for HD~189733b's outer atmosphere given our calculations above, the ratio of H$^+$ to H is ca.\ $45$ ($5$, $0.5$) for temperatures of $23{,}000$~K ($18{,}500$~K, $14{,}000$~K; \citealt{Mazzotta1998}). Consequently, UV measurements of atomic hydrogen should only be sensitive to a small fraction of the total mass loss. In contrast, the X-ray absorption is driven by heavier elements and therefore practically unaffected by hydrogen ionization.

   \begin{figure}[t!]
   \centering
   \includegraphics[width=0.45\textwidth]{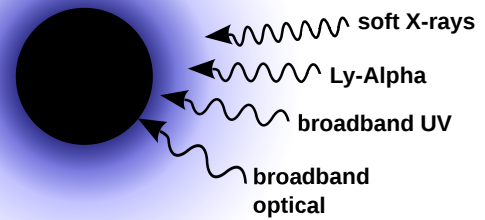}
   \caption{Schematic representation of the absorption altitudes in HD~189733Ab's atmosphere.}
   \label{atmosphere}
   \end{figure}

\subsection{Planetary mass loss due to the extended X-ray radius}

Irradiation of exoplanets in the X-ray and extreme UV regime has been identified as the main driver for their atmospheric escape \citep{Lecavelier2007}. The altitudes at which the high-energy irradiation is absorbed depends on the wavelength. Theoretical models show that the broadband (E)UV flux is absorbed close to the optical planetary surface, at $R_{UV} \approx 1.1 R_{opt}$ \citep{Murray-Clay2009}. In the far-UV hydrogen Ly-$\alpha$ line, the opacity of the atmosphere is much larger, and transit depths between 2.7 and 7.6\% have been inferred from individual observations, with a typical transit depth of ca.\ 5\% \citep{Lecavelier2010}.\footnote{A very large Ly-$\alpha$ transit depth of 14.4\% has been observed following a strong stellar flare \citep{Lecavelier2012}.} Our X-ray observations have shown that the X-ray absorbing radius of the planet is likely even larger with about $1.75 R_P$. This is consistent with our temperature estimate of ca.\ 20,000\,K for the outer atmosphere, which means that hydrogen is mostly ionized and will not contribute to the EUV opacity at those altitudes. We show a qualitative representation of the different absorption altitudes in Fig.~\ref{atmosphere}. Note that the planetary atmosphere is likely distorted by the stellar wind, and may form a comet-like tail \citep{Vidal-Madjar2003}.

Observations in the UV Ly-$\alpha$ line of hydrogen yielded lower limits for mass-loss rates of the order of $10^{10}$\,g\,s$^{-1}$ for transiting Hot Jupiters \citep{Vidal-Madjar2003, Lecavelier2010}. However, the dynamics of planetary mass loss are not well understood; the process is hydrodynamical \citep{Tian2005, Murray-Clay2009} and therefore much more efficient than pure Jeans escape. Current models emphasize different aspects of the problem, such as extended absorption radii Roche-lobe overflow \citep{Lammer2003, Erkaev2007}. For an order-of-magnitude estimate of HD~189733b's mass loss, we choose to use an approximate formula which related the mass loss rate to the stellar high-energy flux, the planetary absorption radius and planetary mass:

\begin{equation}\label{massloss}
\dot{M_P} = \frac{\pi \epsilon F_{XUV} R_P\, R^2_{P\,(abs)}}{ G M_P K},
\end{equation}
with $M_P$ being the planetary mass of $1.138$\,M$_{Jup}$ \citep{Triaud2010}, $R_{P\,(abs)}$ the relevant planetary radius for high-energy absorption, $K$ a factor for including Roche-lobe overflow effects which we ignore here by assuming $K=1$, $F_{XUV}$ the incident high-energy flux at the planetary orbit, and $\epsilon$ a factor to account for heating efficiency of the planetary atmosphere. A choice of $\epsilon=1$ denotes that all incident energy is converted into particle escape, but several authors chose $\epsilon=0.4$ \citep{Valencia2010, Jackson2010}, inspired by observations of the evaporating Hot Jupiter HD~209458b, and we follow their approach. 

Assuming that the broadband EUV flux is absorbed close to the optical planetary radius, while the X-ray flux is absorbed at altitudes of $\approx 1.75 R_P$, we modify formula \label{massloss} to

\begin{equation}\label{massloss2}
\dot{M_P} =\pi \epsilon R_P \frac{(1.75 R_P)^2 F_{X} + R_P^2 F_{EUV}}{ G M_P K}.
\end{equation}

The broadband EUV flux of HD~189733A has not been directly measured, but scaling relations between the EUV and X-ray flux exist for main-sequence stars \citep{Sanz-Forcada2011}, using specific X-ray energy bands. We use WebPIMMS to extrapolate our measured X-ray luminosity of HD~189733A of $\log L_X\,(0.25-2\,keV) = 28.08$ to the required 0.12-2.5\,keV band and find $\log L_{X\,(0.1-2.5)} = 28.15$ for a mean coronal temperature of $4.5$\,MK. This yields an EUV luminosity of $\log L_{EUV} = 29.01$ using formula 3 from \cite{Sanz-Forcada2011}. Another way to estimated the EUV luminosity of HD~189733A is by extending its coronal emission measure distribution, derived from a single high-resolution X-ray spectrum, to EUV wavelengths. \cite{Sanz-Forcada2011} have done this and found $\log L_{EUV} = 28.5$; given that the star displays some X-ray variability over time, we consider a range of EUV luminosties between those two values.

Converting these luminosities to flux at the planetary orbit with a semimajor axis of 0.03142\,AU, we find a mass loss rate between $1.2-2.3\times10^{11}$\,g\,s$^{-1}$ for an energy efficiency of $\epsilon = 0.4$, and a maximum mass loss of $3.0-5.8\times10^{11}$\,g\,s$^{-1}$ for an energy efficiency of $\epsilon = 1$.
The contribution of the extended X-ray radius to the increased mass loss is 25-65\%, depending on the assumed high or low EUV flux which is absorbed at lower altitudes in the planetary atmosphere.

Other studies have tried to estimate the total mass lost by selected exoplanets over the planetary lifetime, by integrating over the age of the system and the likely activity evolution of the host star \citep{Lammer2009, Sanz-Forcada2011, Poppenhaeger2012}. For HD~189733A, however, the activity evolution is probably atypical, as shown in the next section.

\subsection{The age of the HD~189733AB system}

Estimating ages of individual main-sequence stars is a thorny problem. Lithium abundances provide some leverage to derive ages of young stars, but this method loses its diagnostic power for ages larger than ca.\ 600\,Myr. For older stars with outer convective zones, i.e.\ later than spectral type A, one can use the effect of magnetic braking to estimate their ages \citep{Mamajek2008}; older stars rotate more slowly and are less magnetically active than younger stars. This has been successfully calibrated through comparisons of stellar clusters with different ages, using a variety of angular momentum proxies such as X-ray luminosity \citep{Preibisch2005, Lammer2009}, fractional X-ray luminosity $L_X/L_{bol}$ \citep{Jackson2012}, chromospheric Ca\,{\sc ii} H and K line emission \citep{Wright2004chrom}, and rotational periods \citep{Cardini2007}.

All of these age-activity relationships assume that the stellar angular momentum is lost through magnetized stellar winds, and that there is no input of angular momentum through other physical processes. These age-activity relationships are violated by close binary stars because they are tidally locked; their stellar spin is locked to the orbital period and is therefore conserved over a long time, leading to X-ray luminosities which are orders of magnitudes larger than for single stars of comparable age. It has been argued that Hot Jupiters might have similar, but weaker activity-enhancing effects on their host stars \citep{Cuntz2000}. While initial observations found chromospheric and coronal activity enhancements \citep{Shkolnik2005, Kashyap2008}, later studies showed that such trends could not be disentagled from biases inherent to planet detection methods, and thus have to be regarded with care when testing for planet-induced effects \citep{Poppenhaeger2010, Poppenhaeger2011}. It is therefore still under debate how strongly close-in exoplanets can influence the magnetic properties of their host stars.

The HD~189733 system provides a new angle to investigate the stellar activity level, because a physically bound stellar companion is present to which the planet-hosting star can be compared. The distance between the (planet-hosting) K dwarf and its M dwarf companion is large (projected orbital distance $\approx 216$\,AU; \citealt{Bakos2006}), so that the rotational history of each star is not influenced by the other. One would therefore expect both stars to display the appropriate activity level of their spectral class for a common age.

The relevant properties of HD~189733A in this context are $\log L_X = 28.1$\,\ergs, $\log (L_X/L_{bol}) = -5$, chromospheric activity indicator $\log R^\prime_{HK} = -4.537$ \citep{Melo2006}, and the rotational period was measured to be $\approx 12$\,d \citep{Henry2008}. From the X-ray properties we find age estimates of $1.2\,Gyr$ and $1.5$\,Gyr (from \cite{Lammer2009} and \cite{Jackson2012}), the chromospheric activity yields $1.1$\,Gyr (from \cite{Wright2004chrom}), and the rotational period yields $1.7$\,Gyr (from \cite{Cardini2007}). All these estimates agree that HD~189733A displays the activity level of a star with an age around 1.5\,Gyr.

HD~189733B's relevant properties are its X-ray luminosity $\log L_X = 26.67$\,\ergs and its fractional X-ray luminosity $\log (L_X/L_{bol}) = -4.6$. This is in line with the X-ray upper limit previously reported by \cite{Pillitteri2011}. Chromospheric activity indices or the rotational period are not available. Here we find age estimates of $4$\,Gyr and $5.5$\,Gyr, respectively, from \cite{Jackson2012} and \cite{Lammer2009}. This is consistent with the fact that HD~189733B's X-ray luminosity is slightly below the median value found for field M dwarfs by \cite{Schmitt1995}. 

This means that the Hot Jupiter host star HD~189733A is magnetically over-active when compared to its presumable co-eval stellar companion. Interestingly, such a contradiction in activity-derived ages has also been observed for the CoRoT-2 system, where the Hot Jupiter-hosting star is also over-active when compared to its stellar companion \citep{Schroeter2011}. In the case of HD~189733A/B, the system has been observed multiple times over recent years, without large changes in the average X-ray luminosity. We can therefore exclude a stellar activity cycle to be the cause for the disagreement in activity levels. We consider it a more likely possibility that the stellar angular momentum of HD~189733A has been tidally influenced by the Hot Jupiter, which has inhibited the stellar spin-down enough to enable the star to maintain the relatively high magnetic activity we observe today.

\section{Conclusion}

We have analyzed X-ray data from the HD~189733 system, consisting of a K0V star with Hot Jupiter and a stellar companion of spectral type M4. In summary,

\begin{itemize}
  \item{We have detected an exoplanetary transit in the X-rays for the first time, with a detection significance of 99.8\%, using only datasets in which the star was quiescent, and 98.0\% using all datasets;}
  \item{The X-ray data support a transit depth of 6-8\%, compared to an optical transit depth of 2.41\%.  We are able to exclude the presence of a stellar activity belt as the cause for this deep X-ray transit, and also a repeated occultation of the same active region is excluded by the observational cadence. We consider the presence of extended atmospheric layers that are transparent in the optical, but opaque to X-rays as the most likely reason for the excess X-ray transit depth. We note, however, that the available seven observations cannot fully rule out the possibility that a non-typical part of the corona is sampled, which is why additional observations will be necessary;}
  \item{The estimated mass loss rate of HD~189733b is $\approx 1-6\times 10^{11}$\,g\,s$^{-1}$, about 25-65\% larger than it would be if the planetary atmosphere was not extended;}
  \item{We detect the stellar companion in X-rays for the first time. Its X-ray luminosity is $\log L_X = 26.67$\,\ergs, which places the system at a likely age of $4-5.5$\,Gyr, while the primary's activity suggests a younger age of $1-2$\,Gyr. This apparent mismatch may be the result of planetary tidal influences, and remains to be investigated further.}
\end{itemize}

\begin{acknowledgements}
      K.P. acknowledges funding from the German Research Foundation (DFG) via Graduiertenkolleg 1351. This work makes use of data obtained with the {\it Chandra X-ray Observatory} and {\it XMM-Newton}.
\end{acknowledgements}

% \bibliographystyle{apj}
% \bibliography{/home/kpoppen/texmf/katjasbib.bib}

\section{Appendix}

The host star HD~189733A is an active cool star, so that one may expect flares to occur in the stellar corona. If major flaring events are present, the sequence of measured X-ray photons per time bin will not be random, but will display autocorrelation. If autocorrelation is present at significant levels, a more detailed treatment of the errors in the phase-folded light curve has to be performed. We therefore compute the autocorrelation function of each X-ray light curve, using time bins of 200\,s as shown in Fig.~\ref{individual_lcs}, left and top right.

   \begin{figure}[ht!]
   \centering
   \begin{minipage}[ht]{0.45\textwidth}
   \includegraphics[width=1.\textwidth]{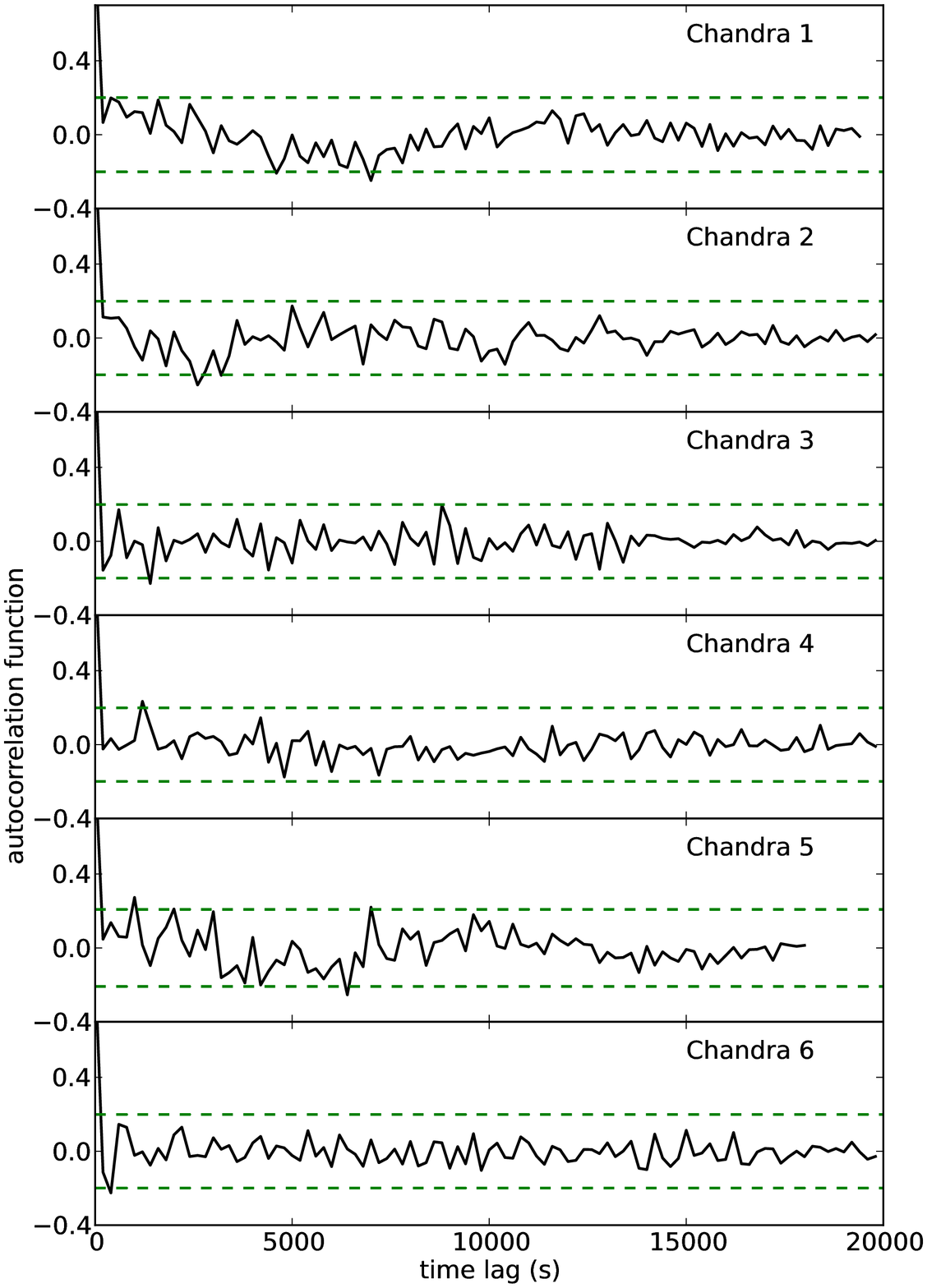}
   \end{minipage}
   \begin{minipage}[ht]{0.45\textwidth}
%    \vspace{-4cm}
   \includegraphics[width=1.\textwidth]{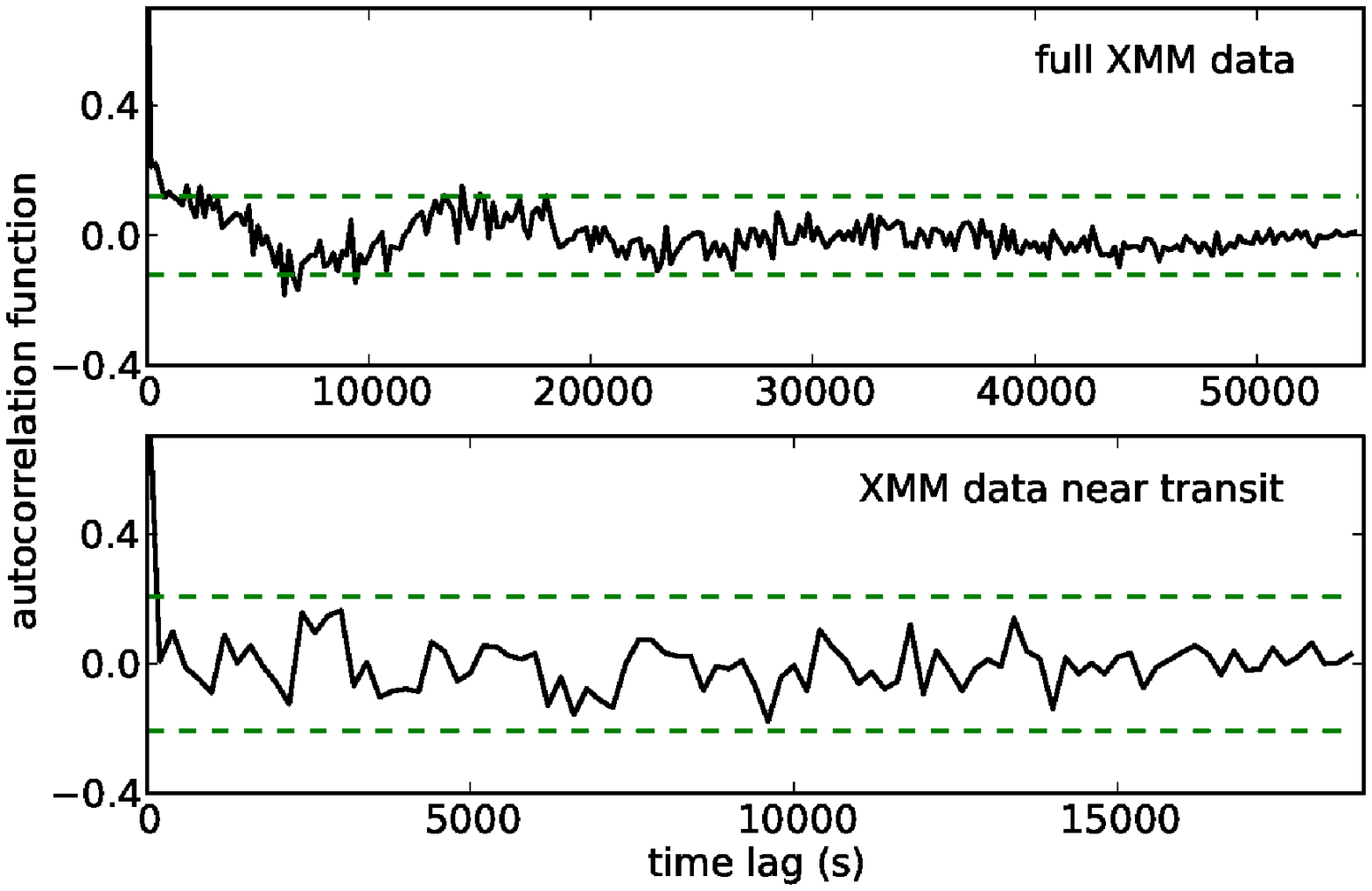}   \vspace{-0.2cm}\\
   \includegraphics[width=1.\textwidth]{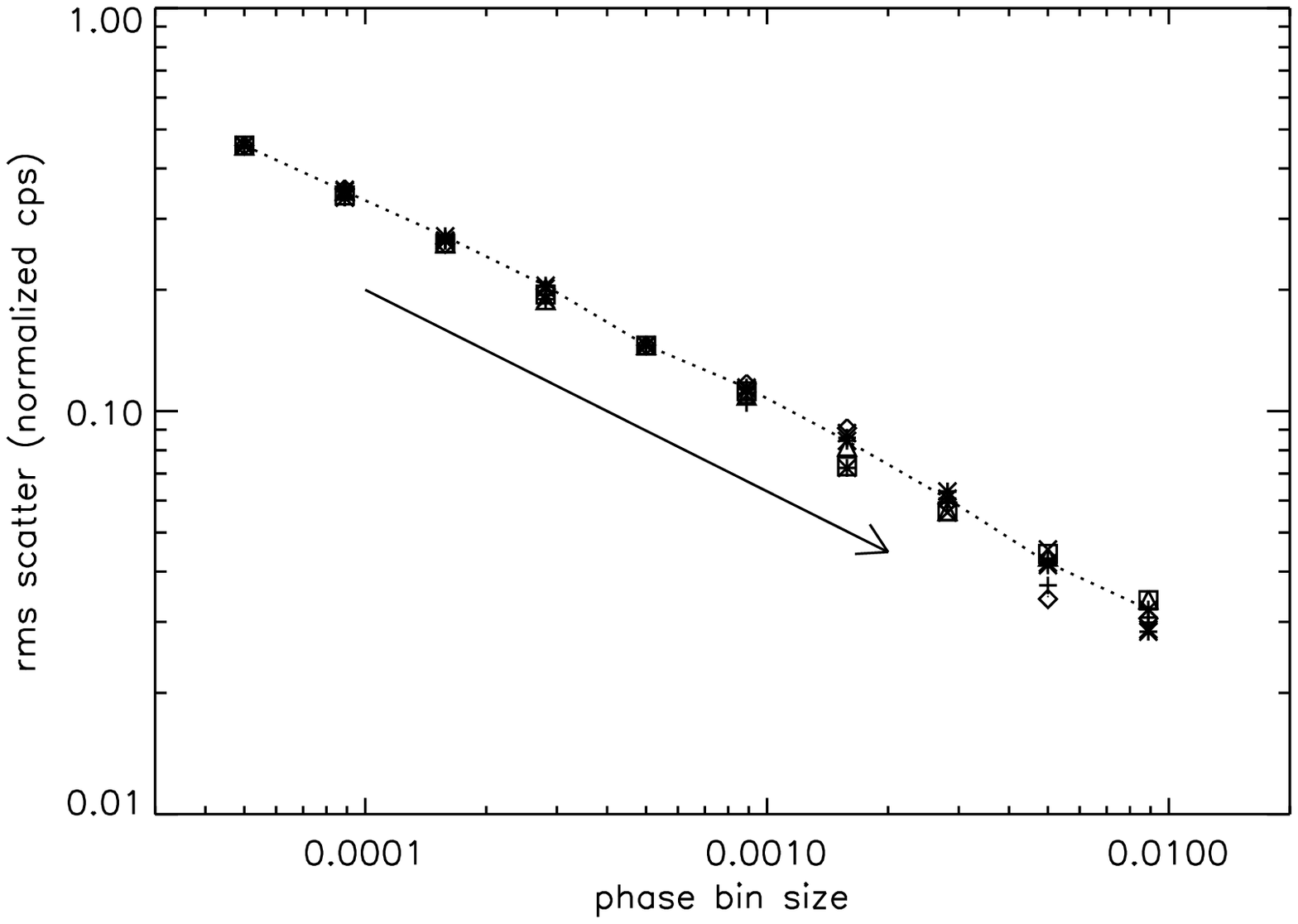}
   \end{minipage}
   \caption{Left: Autocorrelation function for the individual {\em Chandra} light curves binned by 200\,s. Top right:  Autocorrelation function for the {\em XMM-Newton} light curve binned by 200\,s, for the complete length of the observation (top) and for the time interval close to the transit (middle, dotted lines in Fig.~\ref{individual_lcs}). The dashed lines represent the 95\% confidence levels for autocorrelation. Bottom right: The rms scatter of the phase-folded X-ray light curve for different bin sizes. The different symbols depict different binning start points, which cause a spread in scatter at larger bin sizes. The expected slope for white noise is depicted by the arrow. The dotted line depicts the rms of the light curve in Fig.~\ref{XrayLC} at different bin sizes; the final phase binning in Fig.~\ref{XrayLC} is 0.005.}
   \label{autocorr}%
   \end{figure}

We show the result in Fig.~\ref{autocorr} (left) for the Chandra light curves. While there is some slight modulation visible in some of the autocorrelation plots, it does not reach a significant level (95\% confidence threshold depicted by the dashed lines). In the case of the {\em XMM-Newton} observation (Fig.~\ref{autocorr}, top right) which has a longer exposure time, a peak at a time lag of ca.\ 15000\,s is present with a power close to the 95\% confidence level. This corresponds to the two flare-like shapes in the {\em XMM-Newton} light curve (lowest panel in Fig.~\ref{individual_lcs}) at orbital phases 0.87 and 0.94, i.e.\ before the orbital phase segment which is relevant for our transit analysis. The part of the {\em XMM-Newton} light curve which covers the same orbital phase as the {\em Chandra} observations shows no significant autocorrelation (Fig.~\ref{autocorr}, middle right).

In addition to the autocorrelation analysis, we test for red noise, i.e.\ excess scatter at large time bines, in the phase-folded, added-up transit light curve.
We analyzed how the rms (root mean square) scatter of the phase-folded light curve, using all data sets, behaves for varying bin sizes; we used bin sizes ranging from 0.00005 in phase ($\approx$10~s, containing ca. 5 photons) to ca. 0.01 ($\approx$2000~s, the time scale we expect for small to medium flares). For the large bins, we have only a few data points ($\geq 10$). This bears the risk that a single realization of the light curve, with some specific start point for the first bin, may be accidentally very smooth or very noisy. We therefore perform the rms calculation for a variety of bin starting points. The result is shown in Fig.~\ref{autocorr}, bottom right. For white noise, the rms scatter should decrease proportional to $\sqrt{r}$, with $r$ being the ratio of bin sizes. As shown in the figure, the noise in the phase-folded light curve is sufficiently white; there is no increase in rms scatter towards larger bins, which would indicate a significant contribution of correlated noise. In order to not pick an accidentally too smooth binned light curve for the further analysis, we used a bin starting point which has the scatter indicated by the dotted line in Fig.~\ref{autocorr}, bottom right, with 0.005 phase binning.

\end{document}